\documentclass[10pt,journal,letterpaper]{IEEEtran}

\usepackage{ifpdf}
\ifpdf
   \usepackage[pdftex]{graphicx}
  \usepackage{mathrsfs}
  \usepackage{amsmath}
  \usepackage[square, comma, sort&compress, numbers]{natbib}
  \usepackage{algorithm}
  \usepackage{algorithmicx}
  \usepackage{algpseudocode}

  \usepackage{enumerate}
  \usepackage{multirow}

  \usepackage{amsthm}
  \usepackage{color}
  \usepackage[square,numbers,sort&compress]{natbib}
  \newcommand{\tabincell}[2]{\begin{tabular}{@{}#1@{}}#2\end{tabular}}

  \usepackage{array}
  \newcommand{\PreserveBackslash}[1]{\let\temp=\\#1\let\\=\temp}
  \newcolumntype{C}[1]{>{\PreserveBackslash\centering}p{#1}}
  \newcolumntype{R}[1]{>{\PreserveBackslash\raggedleft}p{#1}}
  \newcolumntype{L}[1]{>{\PreserveBackslash\raggedright}p{#1}}

  \usepackage{marvosym}
  \usepackage{CJK}
  \usepackage{amsfonts}
  \usepackage{subfigure}
  \usepackage{url}
  \usepackage{multirow}
\else
\fi

\ifCLASSINFOpdf
\else
\fi

\begin{document}
\title{IriTrack: Liveness Detection Using Irises Tracking for Preventing Face Spoofing Attacks}

\author{
  Meng Shen,~\IEEEmembership{Member,~IEEE},
  Zelin Liao,
  Liehuang Zhu,~\IEEEmembership{Member,~IEEE,}
  Rashid Mijumbi,~\IEEEmembership{Member,~IEEE,} \\
  Xiaojiang Du,~\IEEEmembership{Senior Member,~IEEE,}
  and Jiankun Hu,~\IEEEmembership{Senior Member,~IEEE}
\IEEEcompsocitemizethanks{
%\IEEEcompsocthanksitem This work was supported in part by the National Science Foundation of China under Grant 61602039,
%in part by the Beijing Natural Science Foundation under Grant 4164098, and in part by the China National Key Research and Development Program under Grant 2016YFB0800301.
\IEEEcompsocthanksitem M. Shen, Z. Liao, and L. Zhu are with
School of Computer Science, Beijing Institute of Technology, Beijing 100081, China (email: shenmeng@bit.edu.cn, lzl1918@126.com, liehuangz@bit.edu.cn).
%Prof. Liehuang Zhu is the corresponding author.
\IEEEcompsocthanksitem R. Mijumbi is with the Bell Labs CTO, Nokia, D15Y6NT Dublin, Ireland (email: rashid.mijumbi@nokia.com).
\IEEEcompsocthanksitem X. Du is with the Department of Computer and Information Sciences, Temple University, Philadelphia, PA 19122, USA (email: dxj@ieee.org).
\IEEEcompsocthanksitem J. Hu is with the School of Engineering and IT, University of New South Wales (UNSW), Canberra, ACT 2610 Australia (email: J.Hu@adfa.edu.au).
}
}

\maketitle

\begin{abstract}
  %----------------------------------------------------- Abstract
  Face liveness detection has become a widely used technique with a growing importance in various authentication scenarios to withstand spoofing attacks.
  %which attempt to circumvent face recognition systems by using non-real faces.
  Existing methods that perform liveness detection generally focus on designing intelligent classifiers or customized hardware to differentiate between the image or video samples
  of a real legitimate user and the imitated ones.
  Although effective, they can be resource-consuming and detection results may be sensitive to environmental changes.

  In this paper, we take iris movement as a significant liveness sign and propose a simple and efficient liveness detection system named IriTrack.  %easy-to-implement
  Users are required to move their eyes along with a randomly generated poly-line, and trajectories of irises are then used as evidences for liveness detection.
  IriTrack allows checking liveness by using data collected during user-device interactions.
  We implemented a prototype and conducted extensive experiments to evaluate the performance of the proposed system.
  The results show that IriTrack can fend against spoofing attacks with a moderate and adjustable time overhead.
\end{abstract}


\IEEEpeerreviewmaketitle

\section{Introduction}\label{sec:introduction}
  %-----------------------------------------------------
  \IEEEPARstart{I}{n} recent years, biometric authentication has been widely used as a reliable and convenient way of user identification and access control %\cite{articleJKH, an_introduction_evaluating, ross2008handbook}.
  \cite{articleJKH}.
  Among all types of biometric features (e.g., fingerprint, voice, retina, and palm veins), facial characteristics gain increasing significance as digital images or videos can be easily captured by cameras readily available on smartphones and other mobile devices \cite{thavalengal2016iris}.
  Thus face authentication becomes popular in a wide range of application scenarios.
  Examples include SmartGate developed by the Australian Border Force and the New Zealand customers services for automated border passing \cite{SmartGate}, HSBC's online banking for allowing customs to open a new account using a selfie \cite{HSBC}, and Windows Hello face authentication in Windows 10 for logging in or unlocking one's Microsoft Passport \cite{windows_hello}.
  The popularity of face authentication is also evidenced by the predicted global market growth at a compound annual growth rate (CAGR) of 9.5\% from 2015 to 2022 \cite{FacialReonline}.

  However, a large body of research has demonstrated the vulnerability of face authentication systems under spoof attacks, where an adversary attempts to spoof the authentication system by mimicking facial features of a legitimate user %\cite{chingovska2014biometrics, montecchi2012quantitative}.
  \cite{chingovska2014biometrics}.
  Based on the object used,
  the existing methods for spoofing a face authentication system can be roughly classified into four categories, namely, picture-based attacks, video-based attacks, mask-based attacks, and 2D/3D model attacks.
  For instance, an adversary in photo spoofing attacks can feed a photo of a specific face to a recognition system, while in video spoofing attacks, a video can be presented to provide more sequential information, e.g., environmental changes
and transformations of facial components.

  To defend against spoofing attacks, face liveness detection is proposed to distinguish between the image or video samples of a legitimate on-site user and the imitated ones %\cite{kahm20122d, 7295057, 7310809}.
  \cite{7310809}.
  For instance, when applying for a new bank account,
  the applicant may be required to take specific actions as an evidence of liveness.
  {
  The face authentication system is thus decomposed into two logically independent processes: \emph{face liveness detection} and \emph{face recognition}. Usually, the former is launched to ensure that the image or video samples are provided lively and by the genuine users, while the latter leverages these samples to determine whether the user is authorized.
  In this paper, we focus on the liveness detection process
  and aim at designing efficient solutions.
  }

Face liveness detection has been studied over the past decade %\cite{a_study_of_liveness, 6889816, 7324134}.
\cite{7324134}.
Existing methods can be divided into two main categories according to the features used for drawing conclusions.
The first category mainly focuses on extracting static features from single images to derive differences of environmental features (e.g., textures and light) between the image displaying surfaces and real faces
%\cite{kim2012face, 8304308, boulkenafet2016face, li2004live, 7084662, 6199819, 6671991, drahansky2008experiments, 6612982, lagorio2013liveness}.
\cite{kim2012face, 8304308, boulkenafet2016face, 7084662, 6671991, 6612982, lagorio2013liveness}.
These methods directly capture images and use them as input, which simplifies the procedure of collecting the necessary input data. However, the simplicity of input data makes them sensitive to environmental factors (e.g., illumination and image quality), which can have a severe impact on detection accuracy.
The second category resorts to sequential images or videos to detect changes in environmental features or facial motions so as to match those changes with real situations \cite{8055588, %kim2013face,
tirunagari2015detection, 6961297, lee2013liveness, bao2009liveness, smiatacz2012liveness, %sun2007blinking_based,
 7029052, 4409068, %4341615, 6884911,
 thavalengal2016iris, 7047885, liu2015exploiting}.
These approaches can better fend against spoofing attacks with a high detection accuracy.
However, these approaches usually suffer from high computational and storage complexity as they introduce cumbersome operations, e.g., applying deep learning algorithms on consecutive images.

Inspired by existing studies which demonstrate the effectiveness of performing analyses over eye movements \cite{security_threats_to, rigas2014gaze, liu2015exploiting, articleJKH}, we explore the feasibility of detecting face liveness using iris trajectory caused by intentional eye movements.
Although eye movement is an important sign of liveness,
the following observations make it an extremely challenging task to precisely track the iris for face liveness detection.
First, eye movements, stimulated by user-device interactions, usually introduce significant noise, e.g., an unconscious change of gaze of a user can lead to frequent and unexpected eye movements \cite{galdi2016eye}.
Second, hardware-defined image adjustment strategies vary greatly in cameras and lead to different transformations of captured images, setting barriers to exact comparisons between actual and expected eye trajectories. For instance, a horizontal flip is usually applied to front cameras, making captured eye trajectories reversed.
Finally, complex interaction patterns improve the security in defending against spoof attacks,
but also reduce efficiency due to longer detection duration.
Therefore, the trade-offs among detection accuracy, efficiency, and system security should be carefully studied.

To address these problems, we propose IriTrack, an efficient system to perform liveness detection by tracking iris changes of users.
IriTrack collects iris positions and uses the derived trajectories to draw a conclusion.
It requires no special hardware,
and can therefore be used on any device equipped with a camera and a display.
The main idea of the proposed system is to trade data acquisition complexity for computation complexity,
which can be suitable for many applications.

We conducted experiments to test the sensitivities for eyes to track among different angles with various parameter combinations, by which we balance the trade-off between detection efficiency and accuracy.
Experimental results demonstrate that IriTrack outperforms the state-of-the-art in terms of detection accuracy, with a moderate time overhead. IriTrack is also robust in environmental condition changes, such as light intensity and face-camera distance.

The main contribution in this paper is two-fold:
    \begin{itemize}
      \item We propose IriTrack, a liveness detection system based on eye movement tracking which works on commercial devices with the ability of image capturing and data processing.
          IriTrack achieves computational simplicity and efficiency, without the need for training complex detection models.
      \item We introduce a probability-based random pattern generation method to increase the ability for defending against potential attacks and to balance system performance.
          In order to get rid of the influence of unconscious eye movement on similarity evaluation, we propose a method to compare the skeleton of displayed patterns against collected eye trajectories.
    \end{itemize}

The remainder of this paper is organized as follows:
We present potential spoofing attacks and briefly summarize existing literatures in Section \ref{sec:related_work}.
We describe the basic idea of IriTrack and highlight the challenges in Section \ref{sec:motivation}.
Then, we present the design details of IriTrack in Section \ref{sec:design}, followed by a security analysis in Section \ref{sec:security}.
The implementation of IriTrack in a commercial device is discussed in Section \ref{sec:implementation} and evaluated for efficiency and security in Section \ref{sec:evaluation}.
Finally, we discuss the limitations of the proposed system in Section \ref{sec:discussion} and conclude the paper in Section \ref{sec:conclusion}.
%Introduction

\section{Related Work}\label{sec:related_work}
  %----------------------------------------------------- Related Work
  In this section, we first present typical spoofing attacks that circumvent face recognition systems, and then briefly review existing methods for face liveness detection.

  \subsection{Facial Spoofing Attacks}
    Generally, face recognition systems extract the identity of a face from one or multiple consecutive images.
    A common idea to deceive face recognition systems is to present facial image samples obtained from the intended target user %\cite{a_study_of_liveness, tirunagari2015detection, kose2013on}.
    \cite{tirunagari2015detection, kose2013on}.
    According to the sources from which facial image samples are obtained, facial spoofing attacks can be categorized as follows:

    \textbf{Picture-based attacks.} Displaying face images such as photos or paintings is a convenient way to spoof face recognition systems. An adversary can offer face recognition systems with pictures of the target user to allow required facial features being detected.
	
    \textbf{Video-based attacks.} Similar to pictures, videos are able to expose specific face features. More importantly, videos usually have the ability to provide face recognition systems with necessary sequential information about environmental changes and transformations of facial components.

    \textbf{2D/3D model attacks.} An adversary can build 2D or 3D models of a valid user, which enables transformations of facial components as well as environmental conditions.
    By adjusting animations of each element, these models can be highly customizable.

    \textbf{Mask-based attacks.} To impersonate face features while preserving environmental conditions, another straightforward idea is to equip an adversary with a face mask.
    %{\color{blue} A face mask can be elaborately decorated to camouflage the attackers facial characteristics. Instead, a mask exposes.}
    % \textbf{Mask-based attacks.} To impersonate face features while preserving environmental conditions, a straightforward idea is to equip an adversary with a face mask. {\color{blue} Moreover, the adversary can fabricate masks with holes exposing specific regions of his real face, where both the expected facial features and facial movements can be exhibited.}
    %This can be hard to be detected only by motions or transformations.

 \subsection{Summary of Typical Face Liveness Detection Methods}

    Recently, many face liveness detection methods have been proposed to determine whether image samples are captured from a real user.
    According to the features they use, we can classify them into two main categories, each of which can further be classified into sub-categories, as shown in Table \ref{tab:suammary_related_work}.
    %namely methods using a single image, methods using sequential images, and methods detecting live signs.

    \textbf{Static features.} Static features are referred as features that contain no transformations, or the alterations can all be regarded as extraneous.
    They can be divided into three types: the first two types are \emph{texture features} and \emph{structure features}, which in most cases can be obtained from single images, while the third type is \emph{human physical characteristics}, which can be directly sensed by special hardware.

    Texture features describe the appearance of specific objects and environmental conditions, e.g., the complexity of colour components within faces %\cite{kim2012face, 8304308, boulkenafet2016face, li2004live}.
    \cite{kim2012face, 8304308, boulkenafet2016face}.
    While structure features depict the information of captured images in its composition \cite{7084662, 6671991}.
    For instance, the size of captured faces can be used as a clue for face liveness detection.
    Methods based on static features ignore transformation information in images. Thus, those methods usually take single images as input. Analyses over single image draw conclusions by contrasting differences between real faces and fake faces in shapes and details \cite{kim2012face,8014829,7490042, 8304308, 7986885}, as the displayed surface of a fake face usually exhibits detectable characteristics, e.g., colour differences, variety in image qualities, etc.
    Kim et al. \cite{kim2012face} proposed an approach to distinguish a real live face from a masked face by differentiating both frequency and texture features.
    %The authors carried out a power spectrum based method to analyse frequency features. To analyse the textures of a given image, they applied a description method based on the Local Binary Pattern (LBP), which is one of the most popular techniques for describing texture information of images.
   % The Support Vector Machine (SVM) classifier was utilized for learning detectors with frequency and texture features derived from the input image.
    %The Support Vector Machine (SVM) classifier was utilized for learning detectors with frequency and texture features derived from the power spectrum based method and LBP.
    Dong et al. \cite{8304308} proposed a liveness detection system, which utilizes the gradient of each colour channel in static images to distinguish between real and fake faces.

    These methods are generally computationally inexpensive since they perform analysis only on single images, rather than on videos or sequential images. Moreover, using single images as input reduces the duration of capturing images, ensuring quick response times. However, they might be sensitive to illumination and image quality, as features extracted from single images contain limited information and easily affected by noise. Thus, they can be error-prone and unstable in varying environmental conditions.

    Human physical characteristics are revealed to describe some properties that only a real person could own, e.g., skin temperature and skin resistance \cite{6612982, lagorio2013liveness}.
    To read features of this kind, in most cases, special hardware must be implemented to sense the data of interest.
    Such a detection can be of high accuracy as well as good efficiency since sensors can respond instantly with high precision. However, the hardware requirements would be an obstacle as such sensors bring extra implementation and maintenance cost.
    In addition, these special hardware may not be available on legacy devices.

    \begin{table}
      \centering
      \caption{Summary of typical face liveness detection methods}
      \label{tab:suammary_related_work}
      \begin{minipage}[b]{0.48\textwidth}
        \centering
        \renewcommand\arraystretch{1.3}
        \begin{tabular}{|l|l|l|}\hline
          Category & Sub-category & Samples of Typical Features \\ \hline
          \multirow{4}{*}{\tabincell{l}{Static \\ feature \\ extraction}} & Texture features & \tabincell{l}{
            Local binary pattern \cite{kim2012face} \\
            Colour \cite{8304308, boulkenafet2016face} %\\
            %Fourier feature \cite{li2004live}
          } \\
          \cline{2-3} & Structure features & \tabincell{l}{
            Diffusion speed \cite{7084662} \\
            Image quality \cite{6671991}
          } \\
          \cline{2-3} & \tabincell{l}{
            Physical characteristics
          }  & \tabincell{l}{
            Skin Temperature \cite{6612982} \\
            Face Depth \cite{lagorio2013liveness}
          } \\ \hline
          \multirow{5}{*}{\tabincell{l}{Dynamic \\ feature \\ analysis}} & Texture comparison & \tabincell{l}{
            Texture change \cite{8055588, %kim2013face,
            tirunagari2015detection, 6961297} \\
            Colour entropy \cite{lee2013liveness}
          } \\
          \cline{2-3} & Structure comparison & \tabincell{l}{
            Optical flow \cite{bao2009liveness, smiatacz2012liveness}
          } \\
          \cline{2-3} & \tabincell{l}{Facial motion detection} & \tabincell{l}{
            Eye blinking \cite{4409068} \\ %,sun2007blinking_based
            Pupil dynamics \cite{7029052} \\
            %Lip movement \cite{4341615, 6884911} \\
            Eye movement \cite{thavalengal2016iris, 7047885, liu2015exploiting}
          } \\ \hline
        \end{tabular}
      \end{minipage}
    \end{table}

    \textbf{Dynamic features.}
    %In contrast with static features, dynamic features are mostly extracted from the component transformations of input images.
    Generally, methods based on dynamic feature analyses take videos or sequential images as input, which provide transformation information of environmental and facial components in time series.
    Methods of this category try to make a judgement by matching environmental and facial changes with real situations \cite{8055588, %kim2013face,
    tirunagari2015detection, 6961297, lee2013liveness, bao2009liveness, smiatacz2012liveness, %sun2007blinking_based,
    7029052, 4409068, %4341615, 6884911,
    thavalengal2016iris, 7047885, liu2015exploiting}.
    %Some natural body motions led by nervous reflexes like blinkings of eyes are regularly occurred and can be hard to aware, which can also be used in detection systems.
    %Lin Sun et al. introduce the blinking-based approach for liveness detection using Conditional Random Fields (CRFs) \cite{sun2007blinking_based} which builds a model about eye blinking and compares the model with another discriminative model.

    Czajka et al. \cite{7029052} proposed a solution based on analyses over changes of human irises. The method is based on the fact that human irises would have their size changed in different light intensity levels while printed irises would have no reaction to such changes.
    Chan et al. \cite{8055588} presented a method by computing changes of both facial and environmental textures with and without an extra light source (e.g., a camera's flash).
    They extracted 534 features based on 4 descriptors of faces and background, which are fed to an SVM classifier for liveness detection. The method requires strong stimulation (e.g., flash light) applied directly to user faces, which may affect the user experiences.

    Compared with the methods based on single image analyses,
    methods in this category employ facial and environmental changes, which can better defend against spoofing attacks,
    but also enlarge the detection duration.
    The requirement of input data can increase storage overhead for capturing and saving images.
    The computational complexity is relatively high, as they perform analysis over a series of frames.

    We pay special attention to solutions which take eye actions (e.g., movements and blinking) as a sign of liveness.
    Several methods need to precisely extract eye positions and require special helmet-like hardware or cameras
    \cite{thavalengal2016iris, 7047885}.
    Czajka \cite{7029052} proposed a solution which uses pupil reacts to light changes for liveness detection.
    In order to capture pupil dynamics (i.e., size), it requires changes of environmental light intensity, starting from complete darkness, which may be infeasible in practical usage. Moreover, pupil size can be altered in different psychological states (e.g., stress, relaxation, and drowsiness), leading to degradation of detection accuracy.
    Liu et al. \cite{liu2015exploiting} uses simple and unaltered patterns, making them less reliable in fending against spoofing attacks.

    The system proposed in this paper captures and analyses motions of human eyes for liveness detection.
    %The system allows some inaccuracy when tracking eyes' positions as we perform measurement by extracting skeletons from eyes' trajectories instead of comparing the eyes' trajectories with generated patterns exactly.
    %A probability-based random model is implemented to stop adversaries from successfully making correct predictions.
    Compared with existing methods, IriTrack needs neither pre-computation nor storage of additional data for training classifiers. It is also robust to environmental changes, such as light intensity and face-camera distance.
    %The proposed method analyses motions of human eyes which makes it counteractive to 2D attacks.

    %In the following section, we explicitly depict our idea and present some feasible solutions to fend against spoofing attacks described above.

    %Many of the existing methods are complicated in data analyses which causes the requirement of a relatively high computation consumption. In this work, we design a brand-new method to classify whether a live face is presented or not. We create an environment which allows users to interact with devices by their eyes. The interested data is collected during interactions. Then by analysing the obtained data, we calculate costs and make decisions by individually comparing the costs with a pre-defined threshold.
%Related Work

\section{Motivation}\label{sec:motivation}
  %----------------------------------------------------- Motivation
  %Currently, various authentication systems have already been implemented. A popular one is the pattern unlocking system used by many mobile phones. Inspired by this unlocking system, we propose our liveness detection system named IriTrack. In this section, we introduce how the basic idea of IriTrack different with the idea of the pattern unlocking system. Based on the particular behaviour that IriTrack traces positions of the human eyes and gives results by trajectories, we probe into the feasibility for eyes tracking along with typical angles.
  In this section, we present the basic idea of IriTrack, based on which we then give our reason for extracting similarity by comparing angles and probe into the feasibility for eyes tracking along with typical angles.

\subsection{Basic Idea}\label{sec:basic_idea}
The idea of IriTrack is inspired by the widely used screen lock pattern systems in smartphones,
where lines are drawn by a user over 9 or more dots displayed on the screen and then compared with a pre-defined pattern by an authorized user.
The screen is unlocked if the two patterns are exactly the same.
Similarly, IriTrack can make decision by comparing the trajectory of a user's eye movements with a pre-defined
patten consisting of a certain number of dots and lines.

\begin{figure}[t]
      \begin{minipage}[b]{0.49\textwidth}
        \centering
        \includegraphics[height=2.5cm]{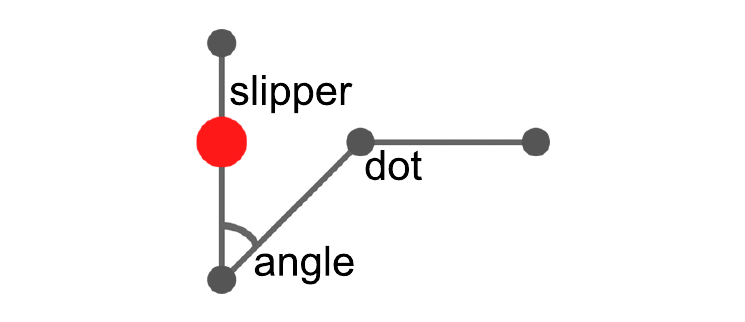}
      \end{minipage}
      \caption{A pattern sample in IriTrack, which contains 4 dots, 3 line segments, and 2 angles. A slipper is employed to direct users' attention between dots.}
      \label{fig:pattern}
\end{figure}

The setting of pre-defined patterns is crucial to the security of a liveness detection system.
A straightforward way is to be consistent with the screen lock pattern systems, where an authorized user can set a customized pattern in advance.
Although simple, it may result in vulnerability as the pre-defined patterns could be leaked to potential attackers.
Additionally, it also imposes the burden of pattern management to users, especially those of different liveness detection applications.
Therefore, we offload the pattern setting operation to the liveness detection system, where a randomly generated pattern is displayed on the screen for a user.

As one has no prior knowledge about the pattern,
it is difficult for a user to determine when to change his attentions.
In order to help users gaze their gaze in an accurate way,
IriTrack uses lines to guide users' attention.
More specifically, in IriTrack, a poly-line with dots inside will be generated and displayed on the screen.
A user has to draw the poly-line by moving his/her eyes.
The trajectory of his/her iris positions is recorded and compared with the given line to get a conclusion.

For clarity, we make several definitions as illustrated in Fig. \ref{fig:pattern}.
In each detection procedure, a \textbf{pattern}, which takes the form of an acyclic \textbf{poly-line} composed of connected line segments, will be randomly generated.
We defer the generation strategy to the next section.
Each pair of adjacent line segments possess an \textbf{angle} seated at their joint. Endpoints of each line segment are referred to as \textbf{dots}.
Correspondingly, eye positions in captured images are called \textbf{points}.

IriTrack differs from the widely used pattern-based screen-lock systems.
Patterns in screen lock systems are pre-defined by users and used as a way of authentication.
However, patterns in IriTrack are \emph{randomly} generated and used only for liveness detection, which is launched before the authentication process performed by face recognition.
The randomness is employed in IriTrack to greatly reduce the possibility of forecasting a pattern by a spoofing attacker.

%{\color{blue}Moreover, as stated in Section \ref{sec:introduction}, we separate liveness detection and face recognition as two processes without interference. As a result, any changes to face images captured by IriTrack are disallowed, preventing the performance of following recognition systems from being affected. }
%During each detection procedure,
%the generated pattern varies in terms of the number of angles, the degree of each angle, and the length of each line segment.
%Second, IriTrack performs analyses in a completely different way, due to differences in locating accuracy between finger touches and irises.

\subsection{Challenges}
It is a non-trivial task to instantiate the above-mentioned idea, due to the following challenges:

\textbf{Unconscious movement of eyes.}
Since IriTrack aims at tracking changes in iris positions caused by users' attention shifts, the fundamental factor affecting the detected result is whether one's eye movements have an anticipant representation.
Existing studies \cite{galdi2016eye} indicate that one's gaze could exhibit unconscious rapid changes, which leads to unexpected eye movements.
What's worse, blinking eyes would also introduce noises in the observed iris trajectory.
%{\color{blue}
%Existing studies by Hoffman \cite{galdi2016eye}
%show that moving one's attention without shifting the gaze is possible, while switching the gaze with attention being held is more difficult.
%In other words, eyes move as the gaze changes in most cases.}
%However, the research also indicates that there are unconscious rapid change of a user's gaze \cite{galdi2016eye}, which leads to not supposed eye movements. Meanwhile, eyes' blinkings would introduce noise in collected trajectories. So, in order to improve the coincidence between eye movements and given patterns, several elements must be involved to help users in guiding their gazes.

\textbf{Transformations of captured images.}
The cameras, operating systems, and hardware in devices can vary greatly due to various manufacturers, which causes the obtained images rendered in different representations.
For instance, a surprising observation in our experiments is that some cameras record images in a horizontally flipped way while others are not.
These uncertain transformations make the exact comparison between eye trajectories and patterns meaningless.
Therefore, we should try to eliminate the impact of such an uncertainty.
%a method to extract the skeleton of a pattern must be proposed to eliminate the impact of such an uncertainty.

\textbf{Trade-offs between efficiency and accuracy.}
As described above, adjusting the number and length of line segments as well as the degree of angles results in various patterns.
Obviously, a longer poly-line with more line segments and angles will prolong the duration of detection,
but also help remove noises in trajectory extraction and thereby improve detection accuracy.
Thus, it is desirable to strike a balance between efficiency and accuracy.
%as the However, a path with long lines inside will stimulate the eyes to move in a larger range, which is certainly easier for similarity measurement in later steps. Thus, there has to be a trade-off between the detection duration and accuracy.

    %Further, We have addressed another problem that may affect the trail of irises' positions. Glasses located between the camera and the human's eyes implicate reflection of additional environment light which should not alter the result of eye region detection but will evidently have influences on searching for the boundary of irises. To allow our system to follow the irises with more accuracy, the reflection should be removed or get thinned.

    \begin{figure}
      \centering
      \begin{minipage}[b]{0.48\textwidth}
            \centering
          \includegraphics[width=0.9\textwidth]{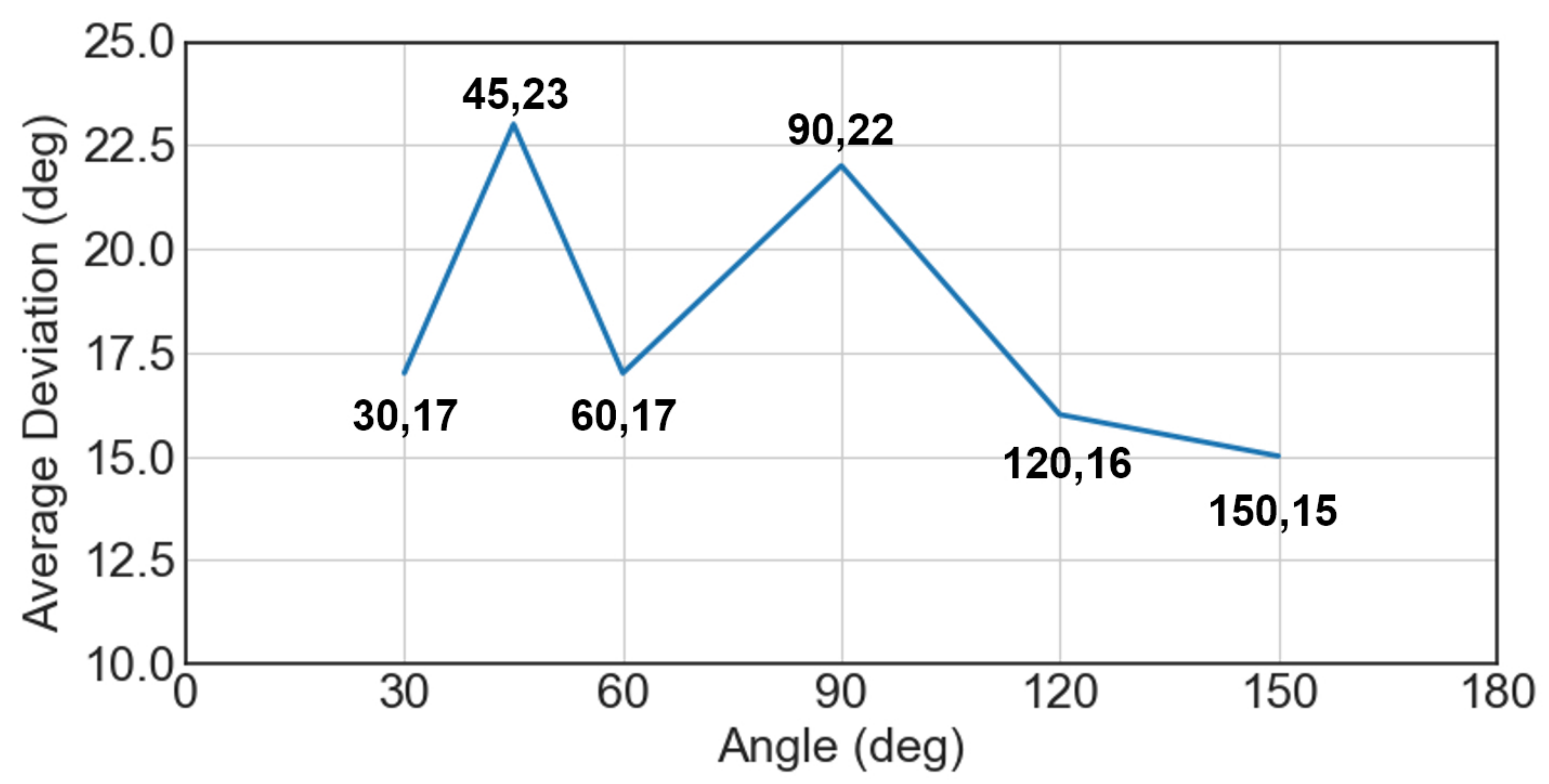}
        \end{minipage}
      \caption{Measured average deviations of angles. A lower deviation means the angle is easier for tracking.}
      \label{fig:sensitivity}
    \end{figure}

\subsection{Sensitivity of Tracking Angles}

In our design, users are required to shift their gaze along with a randomly generated pattern, where the recorded trajectories are then compared with the given poly-lines for making detection conclusion.
However, unconscious eye movements along with eye blinking would result in unpredictable positions of irises, which causes
indeterministic deviation from the poly-lines.
Besides, transformations due to hardware diversity have an influence on the phase of similarity comparison.
Thus, it is extremely difficult to achieve an exact match between the poly-lines and collected trajectories.

In order to address this challenge,
we turn to track eye movements at the critical endpoints in the poly-line.
More specifically,
we view the angles between each pair of adjacent line segments as the skeleton of a pattern,
and attempt to measure the similarity between the skeleton and the eye movements when angles occur.
%Based on our observations, the detection conclusion should be carefully computed or the result can be unexpectedly wrong if it is directly drawn by exactly comparing the collected trajectories with generated patterns.
%Therefore, rather than matching the points sequences with given paths, we turn to evaluating the similarity between eye movements and the skeleton extracted from the poly-line. Ignoring the shape of poly-lines, we take the angles between every two adjacent line segments as the skeleton of a pattern.

To validate the feasibility of the above-mentioned idea,
we conduct experiments to evaluate the sensitivity of tracking eye movements at angles (cf. Section \ref{sec:parameter} for more details).
In the experiments, two lines with an angle at their conjunct endpoint are displayed on the screen, and the positions of pupils are recorded when testers shift their gaze along the given poly-line.
We measure the angle from the tracked irises' positions, and then calculate the deviation of a measured angle from its real value.
Methods for locating iris positions and measuring angles are deferred to the next section.

For the sake of simplicity, we assume that angles on a poly-line are restricted to 6 typical degrees, i.e., $\Gamma = \left \{ 30\rm{deg}, 45\rm{deg}, 60\rm{deg}, 90\rm{deg}, 120\rm{deg}, 150\rm{deg} \right \}$.
Experimental results are shown in Fig. \ref{fig:sensitivity},
where the two numbers at each point indicate the real value of an angle and its deviation, respectively.
From the results, we can learn that it is possible to track eye movements for typical angles.
But the sensitivity varies among different angles, e.g.,
angles of $45\rm{deg}$ and $90\rm{deg}$ are more difficult to track, which should be carefully considered in pattern design.
%Then, we check the deviation of each angle measured from iris trajectories to the corresponding actual value in the first part of our experiment depicted in Section \ref{sec:parameter}.
%The sensitivity for eyes tracking along with a specific angle indicates the easiness that the angle gets followed. We conduct our experiment to measure the sensitivity. Two lines with an angle at their conjunct endpoint is displayed on the screen and the positions of pupils are recorded when the tester tracks along with the given poly-line. We calculate the measured angle from tracked irises' positions, and then, the deviation of a measured angle with the corresponding actual value can be obtained by some subtractions. Methods implemented to locate irises' positions and evaluate angles from trajectories will be explicitly introduced in the next section.
%Part of our experimental result is shown in Fig. \ref{fig:sensitivity}. A lower deviation of an angle means that the angle is easier for eyes to track with.

\section{Design of IriTrack}\label{sec:design}
  %----------------------------------------------------- Problem Formulation
  %IriTrack is overall based on the idea of drawing conclusions from measuring the closeness of irises' motions to a given pattern.
  In this section, we present the workflow and design details of IriTrack.
% According to the process, we give the detailed design of modules that make up our system.
%Considering potential spoofing attacks mentioned previously, we give some analyses on its security when shielding against those attacks.

\subsection{System Overview}

We build our system based on two primary facts \cite{galdi2016eye}.
First, one can keep staring at a specific object for a relatively short time (e.g., 5 seconds).
Second, tracking any specific object with eyes causes detectable changes of relative distances between eye regions and the center of irises in the prerequisite of keeping one's head still.

\begin{figure}
      \centering
      \begin{minipage}[b]{0.48\textwidth}
          \includegraphics[width=1\textwidth]{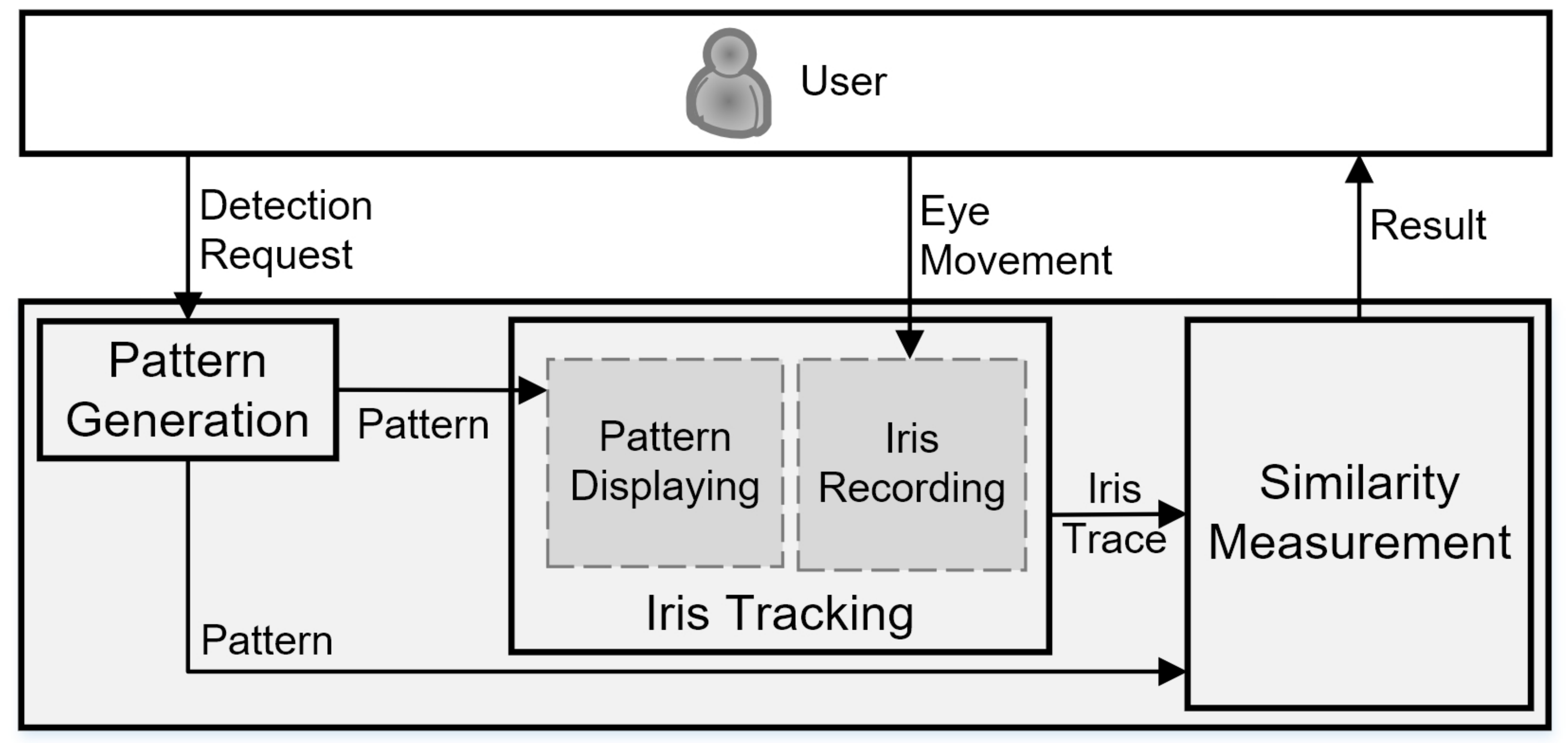}
        \end{minipage}
      \caption{The workflow of IriTrack.
      A random pattern is displayed on the screen when receiving a request from a user, and the user is required to draw the pattern with his/her eyes. Then, the recorded iris trajectories along with the original pattern are taken as inputs for similarity measurement. Finally, the detection result (i.e., pass or fail) is returned to the user.}
      \label{fig:workflow}
\end{figure}

IriTrack's system architecture is illustrated in Fig. \ref{fig:workflow}, and is mainly composed of three components, namely pattern generation, iris tracking, and similarity measurement.
The design details of each module will be described in
the following subsections.

The workflow of liveness detection can be described as follows: IriTrack randomly generates and displays a pattern on the screen as requested by a user.
Then, the user is required to follow the pattern with his eyes
and the trajectories of irises can be recorded by a camera. During this process, the user is required to try his best to keeping his head still.
Finally, the collected trajectories and the given pattern are fed to the similarity measurement module for drawing a conclusion.

{
Since liveness detection samples must be used for recognition as well, the liveness face images, from which iris patterns are retrieved in IriTrack, do not affect the recognition by certain cutting-edge recognition algorithms if used as recognition input. As face recognition is logically independent of liveness detection, we focus on the design details of IriTrack hereafter.}

  \subsection{Pattern Generation}
    As stated, a pattern is a poly-line consisting of line segments. To help users concentrate and balance their tracking speed, a slipper which moves along with patterns at a constant speed is also displayed. All patterns need to be arbitrarily generated to avoid potential spoofing attacks.
    %At the same time, as the poly-line gets longer in its length, the strength of IriTrack preventing against attacks would be improved, while the time for users' eyes to draw the poly-line would become longer.
    In our design, we take consideration of the following two factors in pattern generation: 1) the capability in fending against spoofing attacks, and 2) time efficiency of tracking iris positions.

    A pattern $\mathcal{P}$ is denoted by $\mathcal{P}=(\mathcal{A}, \mathcal{L})$,
    where $\mathcal{A}$ and $\mathcal{L}$ are angle set and line segment set in the pattern, respectively.
    The generated patterns should be random enough, or attackers may take preparations in advance if a pattern can be easily speculated.
    To allow the randomness of patterns, we apply probabilities
    when generating angles and lines.
    Recall that $\Gamma$ contains typical angles from which an angle in a pattern $\mathcal{P}$ can be selected.
    For each angle $\theta \in \Gamma$, we associate a weight $\omega_\theta$, which indicates the probability of accurately following such an angle by eye movements.
    The notations used in the rest of this paper are summarized in Table \ref{tab:notations}.
     %As the weight of an angle describes the accuracy of irises following along with it, an angle with a higher weight should be more possible to occur in a pattern.

    We denote $P_\Gamma(\theta)$ as the probability of setting the current angle to be $\theta$ ($\theta \in \Gamma$).
    $P_\Gamma(\theta)$ can be calculated as follows:
    \begin{equation}\label{eq:pa}
      P_\Gamma(\theta)=\frac{\omega_\theta}{\sum_{\theta_i \in \Gamma} {\omega_{\theta_i}}}
    \end{equation}
    where $\Gamma = \left \{ 30\rm{deg}, 45\rm{deg}, 60\rm{deg}, 90\rm{deg}, 120\rm{deg}, 150\rm{deg} \right \}$.

    It can be noted that the higher the number of dots used, the more difficult a spoofing attack succeeds.
    Let $n$ be the total number of dots in a generated pattern.
    Here, we assume $n \geq 4$ (i.e., at least two angles in a pattern) for security considerations.
    There would be $6^{n-2}$ possible combinations of angles.
    Meanwhile, line segments appearing in a pattern are randomly selected from a pre-defined set $L= \left \{ l_0, l_1, \ldots, l_{\left | L \right |} \right\}$,
    thus there would be ${\left | L \right |}^{n-1}$ possible combinations of line segments.
    %Theoretically, the attacker should prepare at least $6^{n-2}${\color{blue}$\times 3^{n-1}$} slices of videos and pick a correct one within a short time to cheat the system. As a result, the increasing number of dots introduces extra security into our system.
    It should be noted that increasing the number of dots can cause increment of the time spent gathering iris tracks, and also make users impatient, which can affect the accuracy of tracking.
    As the slipper moves at a constant speed $s$,
    the time cost should be directly proportional to $n$.

    To achieve a balance between time cost, security against spoofing attacks, and tracking accuracy,
    we resort to a probability-based model of pattern generation,
    where we start from a pattern $\mathcal{P}$ with only one line segment (i.e., two dots), and iteratively determines whether a new line (also a new angle) should be added to the current pattern, as stated in Algorithm \ref{algorithm:PatternGeneration}.

    In each iteration, assume that $k-1$ ($k \geq 3$) dots already exist in pattern $\mathcal{P}$,
    we use $P_L(k)$ to denote the probability of adding the $k$-th dot in $\mathcal{P}$, as shown in Eq. \eqref{eq:pl}. % {\color{blue} Our experiments (cf. Sec. \ref{sec:scale_pattern}) show that once more angles are involved, the difficulty of succeeding in spoofing a pattern increases, while the time overhead augments in the same time. To achieve a balance, we let each pattern have at least 3 angles (or 5 dots), and restrict the probability of introducing more angles at a low level.}
    Algorithm \ref{algorithm:IsNextDotNeeded} shows how to determine whether a dot should be added.
    %Once the judgement returns false, a pattern is regarded to be successfully generated.
    \begin{equation}\label{eq:pl}
      P_L(k) = \left \{
        \begin{array}{lr}
          1, & k \leq 4 \\
          \frac{1}{k-3}, & k > 4
        \end{array}
        \right.
    \end{equation}

\begin{table}[t]
  \caption{Notations used in IriTrack}
  \label{tab:notations}
  \centering
  \renewcommand\arraystretch{1.3}
  \begin{tabular}{|l|l|}\hline
    Notation & Description \\  \hline
    $\Gamma$ & The set containing typical angle degrees \\ \hline
    $L$ & The set containing typical lengths of line segments \\ \hline
    $\mathcal{P}$ & A generated pattern \\ \hline
    %where $\mathcal{P}=\left ( \mathcal{A}, \mathcal{L} \right ) $
    $\mathcal{A}$ & The set containing angles in $\mathcal{P}$\\ \hline
    $\mathcal{L}$ & The set containing line segments in $\mathcal{P}$ \\ \hline
    $l$ & The total length of line segments in $\mathcal{P}$ \\ \hline
    $n$ & The total amount of dots in $\mathcal{P}$ \\ \hline
    $s$ & The constant moving speed of the slipper in $\mathcal{P}$ \\ \hline
    $\mathcal{C}$ & The set containing sequentially recorded eye positions \\ \hline
    $P_\Gamma(\theta)$ & The probability of setting the next angle as $\theta$ \\ \hline
    $P_L(k)$ & The probability of adding the $k$-th dot into a pattern \\ \hline
  \end{tabular}
\end{table}

    \begin{algorithm}[t]
      \caption{IsNextDotNeeded}
      \label{algorithm:IsNextDotNeeded}
      \begin{algorithmic}[1]
        \Require $k$ as the index of the next dot to be generated
        \Ensure whether a new dot should be added to the pattern
        \If{$n < 3$}
          \State \Return $true$
        \Else
          \State $prob \gets P_L(k)$
          \State $rand \gets \rm{generate\ a\ random\ number\ between\ 0\ and\ 1}$
          \State \Return $rand \le prob$
        \EndIf
      \end{algorithmic}
    \end{algorithm}

Next, considering the two key factors mentioned in the beginning of this section,
the goodness of the generated pattern should be measured to ensure that the pattern is secure enough to resist against spoofing attacks and requires moderate tracking time.
Given a generated pattern $\mathcal{P}$, we use $G(\mathcal{P})$ to describe its goodness,
which is calculated in Eq. \eqref{eq:goodness}.
    \begin{equation}\label{eq:goodness}
        G(\mathcal{P}) = 6^{n-2} \times {\left | L \right |}^{n-1} \times \frac{1}{e^{n-1}} \prod_{\theta_i \in \mathcal{A}}{P_\Gamma(\theta_i)}
    \end{equation}

The coefficient $6^{n-2} \times {\left | L \right |}^{n-1}$ is a measure of the randomness of $\mathcal{P}$ which is directly associated with the strength in fending against spoofing attacks. The denominator $e^{n-1}$ ensures that a pattern with less line segments is more likely to be accepted as the time overhead for tracking can be reduced. Additionally, we employ an exponential function to introduce a rapid drop in goodness when the time overhead increases. The rest signifies the efficiency for eyes to track angles. A better pattern $\mathcal{P}$ should have a higher value of $G(\mathcal{P})$. A pre-defined constant $G_0$ is introduced and each valid pattern $\mathcal{P}$ must satisfy the condition $G(\mathcal{P}) \ge G_0$. The setting of $G_0$ will be described in Section \ref{sec:evaluation}.

    Moreover, as the pattern would be displayed in the screen, it should be guaranteed that all dots are placed within the bound of the screen. Meanwhile, to reduce the confusion for users when tracking, we stipulate artificially that all lines and dots in a pattern are not allowed to overlap.

    Finally, the above conditions are considered together to
    determines whether to return the current pattern or generate a new pattern (Line 9 in Algorithm \ref{algorithm:PatternGeneration}).
    %Finally, we present the algorithm that the module uses to generate patterns. As described in Algorithm \ref{algorithm:PatternGeneration}, a pattern is initialized with two dots, which contains no angle and one line segment with randomly selected value in its length. In each iteration, only one angle and one line are selected by probabilities and added to the pattern if the next dot needs to be appended. Otherwise, the algorithm calculates the pattern's goodness and determines whether to return the current pattern or generate another one.

    \begin{algorithm}
      \caption{PatternGeneration}
      \label{algorithm:PatternGeneration}
      \begin{algorithmic}[1]
        \Require $\Gamma, L$
        \Ensure Pattern $\mathcal{P}$
        \While{true}
        \State $\mathcal{A} \gets \left \{ \right \}$
        \State $\mathcal{L} \gets \left \{ \rm{a\ randomly\ selected\ line\ from}\ \emph{L} \right \}$
        \Statex \ \ \ \ // There are 2 dots in the current pattern.
        \Statex \ \ \ \ // Index of the next dot would be $\left | \mathcal{A} \right | + 3$.
        \While{$\rm{IsNextDotNeeded}(\left | \mathcal{A} \right | + 3)$}
          \State $\mathcal{A} \gets \mathcal{A} \cup \left \{ \rm{select\ an\ angle\ from}\ \emph{$\Gamma$} \right \}$
          \State $\mathcal{L} \gets \mathcal{L} \cup \left \{ \rm{select\ a\ line\ from}\ \emph{$L$} \right \}$
        \EndWhile
        \State $\mathcal{P} \gets (\mathcal{A}, \mathcal{L})$
          \If{$G(\mathcal{P}) \ge G_0$ \textbf{and}
          \Statex    \ \ \ \ \ \ \ $\rm{all\ dots\ of\ \mathcal{P}\ seat\ within\ the\ screen}$ \textbf{and}
          \Statex    \ \ \ \ \ \ \ $\rm{no\ overlap\ between\ dots\ and\ lines}$ }
            \State \Return $\mathcal{P}$
          \EndIf
        \EndWhile
      \end{algorithmic}
    \end{algorithm}
  %Pattern Generation

  \subsection{Iris Tracking}
    The tracking module utilizes the embedded camera to grab facial images, which are used to identify the center of each iris and track the movements of irises.

    As the module starts working, the camera acquires images at a fixed frequency.
    %Once an image is recorded, the module tries to find the center of irises from the facial image.
    Given a facial image, the Daugman's integrodifferential operator \cite{iris_localization} is employed to detect the center of irises.
    To find a circular path that fixes the contour of each iris, the algorithm tries every combination of center position $(x, y)$ and radius $r$ to detect the path with the maximum change of pixel values.
    It can be expressed by the following equation:
    \begin{equation}
      \max_{r, x_0, y_0}{\left | G_\sigma(r) \ast \frac{\partial}{\partial r} \oint {\frac{I(x,y)}{2\pi r} ds} \right |}
    \end{equation}
    where $I$ is the input image, $I(x,y)$ is the pixel value in the corresponding position $(x, y)$, $r$ is the radius of the detected area, and $G_\sigma(r)$ is the Gaussian smoothing function.

    As we need only transformations of iris positions,
    the coordinates with values of $x_0$ and $y_0$ are recorded, but the detected radius of each iris $r$ is simply ignored.

\subsection{Similarity Measurement}
    %As stated, checking liveness by matching exactly between collected trajectories and generated patterns is not ideal as the image capture devices may apply transformations to the images. As a result, angles inside patterns, a.k.a the skeleton of given patterns, are preferred.  More exactly, in our system, line segments in a pattern are various in length, and the internal angles are generated randomly. The measurement module would try to recover the angles from these clues. Based on the assumption that gaze of eyes moves in a uniformed speed, the coordinates of tracked dots can be proportionally divided according to the length of each line segment.
    As shown in Fig. \ref{fig:workflow}, a randomly generated pattern along with the collected eye trajectories would be passed to our measurement module. The main task of this stage is to recover the skeleton from eye movements and compare the similarity between the skeleton with the given pattern.

    Based on the assumption that the gaze of eyes moves at a uniform speed, the coordinates of tracked dots can be proportionally divided according to the length of each line segment.
    Given a pattern $\mathcal{P}=(\mathcal{A},\mathcal{L})$, let $\mathcal{L}=\left \{ l_1, l_2, \ldots, l_{\left | \mathcal{L} \right |} \right \}$, where $l_v$ ($1\le v \le \left | \mathcal{L} \right |$) is the length of the $v$-th line in $\mathcal{P}$.
    The total length of the poly-line can be denoted by $l=\sum_{l_v\in\mathcal{L}} {l_v}$.
    Denoting $\mathcal{C}=\left \{ \mathcal{C}_1, \mathcal{C}_2, \ldots, \mathcal{C}_m \right \}$ as the set of recorded dots where $m=\left | \mathcal{C} \right |$, the position of the $i$-th dot in the pattern can be recovered as follows:
    \begin{equation}
      \label{eq:divide_track}
      (x_i, y_i) = \left \{
        \begin{array}{lr}
          \mathcal{C}_1, & i = 1 \\
          \mathcal{C}_m, & i = n \\
          \mathcal{C}_j, & otherwise
        \end{array}
        \right. \\
    \end{equation}
   %where $j=\frac{\sum_{v<i}{l_v}}{l} \times m$.
   where $j=\frac{m}{l} \sum_{v<i}l_v$.
   Sequentially taking three dots recovered from the tracked points, the angles in degrees can be easily obtained using the law of Cosines.
   According to Eq. \eqref{eq:divide_track}, distances between the adjacent dots are the rule to recover relative positions among dots, while angles calculated consecutively are the evidence to judge whether the movement of irises are similar to the given pattern.

   {
   In IriTrack, the PC screen and camera captured images usually have two different coordinate systems.
   Since we use angles for similarity measurement, the calculation involved is irrelevant to the coordinate systems.}

   As mentioned above, we assign weights to different angles.
   An angle with a higher weight can be followed with less disparity, and the difference between the angle and its measured value can be more credible.
   We introduce the matching cost $C$ to describe the dissimilarity between the original pattern and the tracked trajectory, as shown in Eq. \eqref{eq:cost}:
    \begin{equation}
      C=\frac{1}{\sum_{\theta_i \in \mathcal{A}} {\omega_{\theta_i}}} \times \sum_{\theta_i \in \mathcal{A}}({\omega_{\theta_i} \times \left | \theta_i - \theta_i^{'} \right |})
      \label{eq:cost}
    \end{equation}
    where $\theta_i $ is the actual value of an angle in the given pattern, and $\theta _i^{'}$ represents its measured result.
    A pre-defined constant threshold $C_{0}$ is involved.
    If $C \leq C_{0}$, we consider that the face in front of the camera comes from a live person.

\section{Security Analysis}\label{sec:security}

  As described above, IriTrack uses eye movements as the evidence for determining the liveness of a presented face.
  In this section, we discuss the security guarantees provided by IriTrack against the potential attacks presented in Section \ref{sec:related_work}.

  %So, any attack, no matter what techniques it utilizes, must allow IriTrack to trace the center of irises and the collected trajectories must be similar to given patterns. Based on this, we come up with some solutions for fending against the following attacks.

  %Considering existing attacks \cite{a_study_of_liveness, tirunagari2015detection, kose2013on}
	
  \textbf{Picture-based attacks.}
  Faces recorded by pictures (e.g., photos) are inherently different from real faces, because the irises in pictures are static..
  As a result, to cheat IriTrack, an attacker must move the picture along with a same path as the displayed poly-line.
  However, this would result in a relatively large range of face movement.
  By analysing the region of face movements during this process,
  IriTrack can easily figure out that the trajectory is derived mainly from face movements, rather than iris movements.

  \textbf{Video-based attacks.} Videos recording eye movements may be used to deceive IriTrack.
  In order to succeed in passing the verification,
  an attacker should present a video displaying a series of eye movements which match the generated poly-line.
  As the poly-line is generated with a high degree of randomness (e.g., the length of segments and the degree of angles), it is difficult to spoof IriTrack without a prior knowledge of the displayed ploy-line.
  Experimental results will be presented in Section \ref{sec:evaluation}.
	
  \textbf{2D/3D model attacks.} Although a model can have moveable facial components,
  changing the movements of facial parts usually needs time-consuming reprogramming.
  Thus, a time-out rule can be involved to prevent programming operations. That is, IriTrack can trigger a time-out rule and terminate the detection process with a rejection once the tracking module fails to record eye movement within a certain period.
  %{\color{blue}Furthermore, it can be also required that the target face should be able to be detected from the beginning to the end during the whole verification process.}
  %Also, because those models can only be shown by a screen, those anti-spoof techniques that analyse facial textures described earlier can be adopted to detect this kind of attacks.
	
  \textbf{Mask-based attacks.} Masks of faces expose specific facial features to IriTrack. Similar to pictures, masks are not able to provide irises transformations as eyes within masks are not moveable. Thus, the same idea of detecting pictures attacks can be applied.
  As a variation of mask-based attacks, an adversary may use a mask which have some level of transparency around eyes such that a camera still detect iris movements.
  We will discuss this special case in Section \ref{sec:discussion}.
  %An attacker wearing a mask has the ability to provide transformation information required by IriTrack,
  %{\color{blue}which is hard to detect only by judging in a dynamic point.} In this case, existing methods analysing textures of faces, e.g., the texture-based method \cite{li2004live}, can be combined to avoid this kind of attacks.
	
  %Iris Measurement

\section{Implementation}\label{sec:implementation}
  We have implemented a prototype of IriTrack on a PC with Windows 10.
  This section presents the implementation details.

  During pattern generation, we use a pseudo-random number generator to simulate probabilities.
  %Noticing that a pseudo-random number follows a predictable fashion with deterministic mathematical formulas,
  An alternative way is to obtain random number generator via RANDOM.ORG \cite{random_org_analysis}.
  %In order to provide stronger security guarantees,
  %an alternative is to acquire random numbers by accessing the APIs provided by RANDOM.ORG.
  %Compared with pseudo-random generators, RANDOM.ORG produces random numbers based on atmospheric noise, which are statistically closer to a uniformed distribution \cite{random_org_analysis} and thereby less divinable. Performance of {\color{blue}our framework with these two strategies} is evaluated in Section \ref{sec:evaluation}.

\begin{figure}
      \centering
      \subfigure[Captured raw image] {
        \begin{minipage}[b]{0.2\textwidth}
          \includegraphics[width=0.8\textwidth]{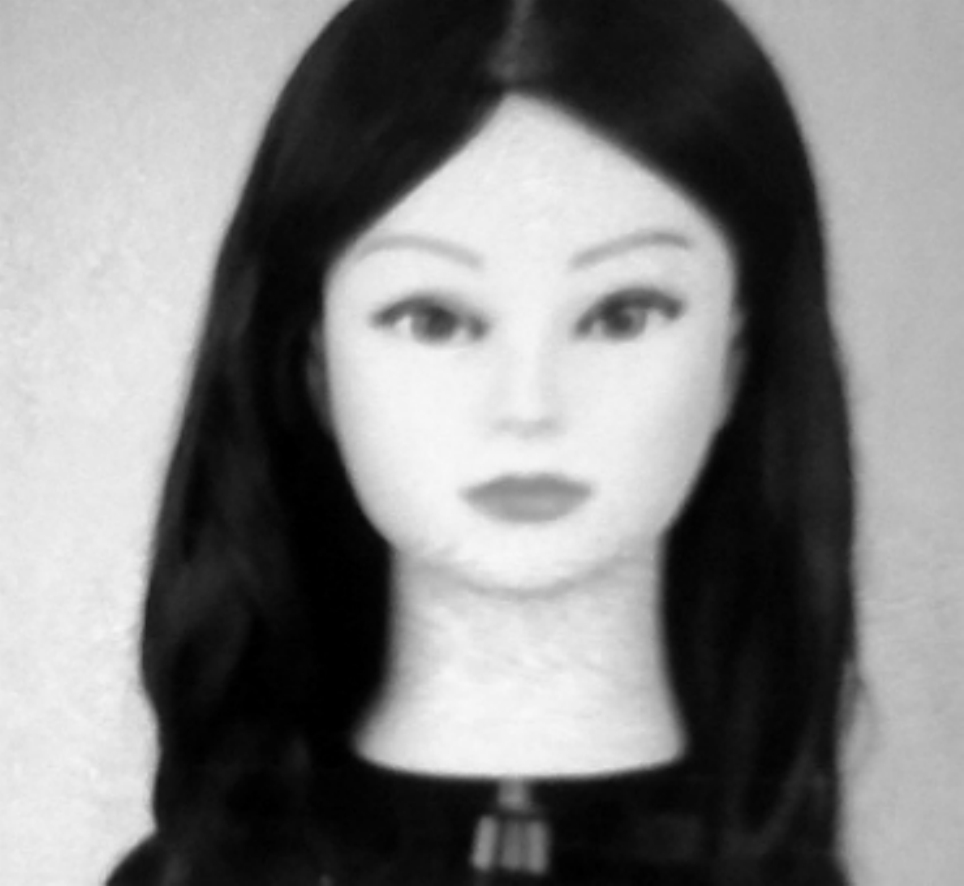}
        \end{minipage}
      }
      \qquad
      \subfigure[Image with detected regions] {
        \begin{minipage}[b]{0.2\textwidth}
          \includegraphics[width=0.8\textwidth]{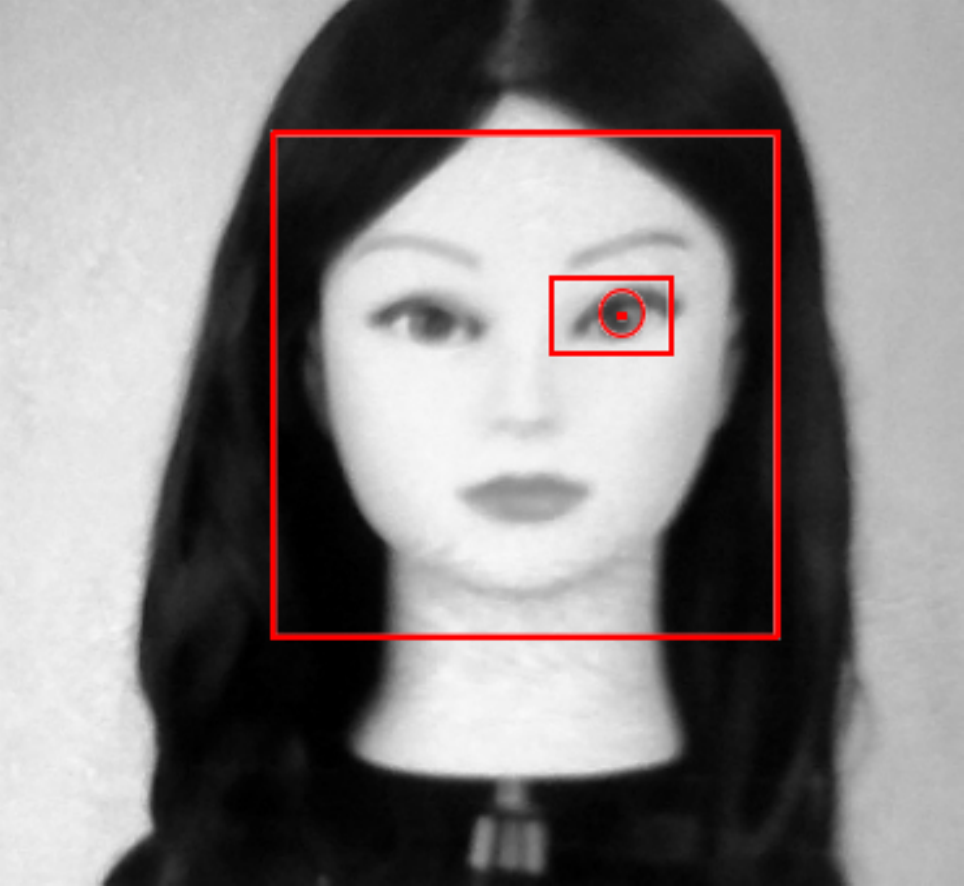}
        \end{minipage}
      }
      \caption{Sample of locating the right iris in a face.
      The region of a user's face is first detected, within which the region of right eye then gets extracted. Finally, the center of right iris could be obtained.}
      \label{fig:track_sample}
    \end{figure}

  The tracking module utilizes OpenCV to invoke image-related functions, e.g., recognizing regions of faces and eyes.
  We use pre-trained Haar classifiers to search for regions of the largest face as well as both the left and right eyes. With the help of the eye classifier,
  IriTrack can successfully detect regions of eyes either with or without glasses.
  By limiting search within the regions of eyes, the locating algorithm is greatly accelerated.
  Fig. \ref{fig:track_sample} demonstrates the result of recognizing regions of interest within a captured face.

    %The Daugman's integrodifferential operator searches the circular path which shares the maximum change of pixel values by individually generating all potential value combinations of circle radius $r$ and the center position $(x,y)$ of the path within the detected regions of eyes. Since each eye region contains only one ring which stands for the iris inside, the operator can successfully find the position of users' irises.

  As stated earlier, IriTrack is supposed to capture points at a fixed frequency.
  When extracting angles from the tracked points, the captured points can be proportionally divided based on the lengths of line segments.
  In other words, the position of a given dot can be derived by the point in the corresponding index from the obtained point sequence as illustrated by Eq. \eqref{eq:divide_track}.
  In our implementation, for each dot in a pattern, we select the corresponding point as well as the 2 nearby points.
  That is, we take 3 points for each dot as its candidates. We maintain those selected candidate sets in a list, from which we sequentially take 3 adjacent sets to calculate angles. For 27 combinations of coordinates respectively selected from the 3 sets, we can get the containing angle by applying the arc-cosine function.
  Finally, we can simply select the most frequent value from the 27 candidates as the final result.

  However, in our experiment, we notice that irises' positions may not be strictly periodically recorded as the processing time may differ for each frame, especially in the situation where some irrelevant background tasks are executed concurrently in the host device.
  The difference between sampling intervals may cause a significant effect in positioning interested dots as well as measuring angles.
  As a result, some revisions must be applied to fix the inaccuracy caused by the uneven scatter of analyses in time series. In our system, we use the captured timestamp of two adjacent points to predicate the position of the user's gaze in a specific moment. When tracking the position of the user's irises, the tracking module is designed to record the $x$ and $y$ coordinates as well as the timestamp when the currently analysed frame captured. We denote $(t_M, (x_M, y_M))$ to be the extracted information from a center point $M$ where $t_M$ represents the time when the position of $M$ is concluded. Having the corresponding data of point $M$ and $N$, we can predicate the position of a point $O$, which is supposed to be recorded at a specific moment $t_O$.
    \begin{equation}
      \label{eq:correction_with_timestamp}
      \begin{aligned}
        x_O = x_M + \frac{t_O-t_M}{t_N-t_M}\times (x_N-x_M) \\
        y_O = y_M + \frac{t_O-t_M}{t_N-t_M}\times (y_N-y_M)
      \end{aligned}
    \end{equation}

  By introducing the timestamp based correction, we can then recover the turning points in a more precise way. We subtract the timestamp of the last recorded point from the timestamp of the first point to extract the duration of the whole process so we can divide the time gap according to the lengths of generated line segments to get the recording moment of turning points.
  %The improvement in detection accuracy will be evaluated in the next section.

\section{Evaluation}\label{sec:evaluation}

  The goals of our evaluation are: 1) exploring parameters that achieve a balance between time overhead and accuracy of the detection process, 2) showing the efficiency and security of our system by comparing with state-of-the-art methods,
  3) demonstrating the system performance with various pattern scales, and 4) estimating the reliability of the proposed system under varying environmental conditions.
  %The following content expands on the above goals.

  \subsection{Preliminaries}
    \textbf{Methods to Compare.} We select several representing liveness detection systems for performance comparison, which are listed as follows:
      \begin{itemize}
        \item IriTrack, which is the main work of this paper. The timestamp-based optimization is involved in similarity measurement.
        \item ncIriTrack, which is the same as IriTrack except the timestamp-based optimization.
        %\item rnIriTrack, which is a variation of IriTrack where random numbers are obtained from RANDOM.ORG via network.
        \item FlashSys, which is the flash-related face liveness detection system proposed by Chan et al. \cite{8055588}.
        \item OptFlowSys, which is proposed by Bao et al. \cite{bao2009liveness} to detect face liveness based on the optical flow field.
      \end{itemize}

    %Here, rnIriTrack is employed to evaluate the time efficiency of different generators of random numbers, while ncIriTrack, FlashSys, and OptFlowSys are involved in the evaluation of detection accuracy.

    \textbf{Testbed.}
    {
    The system is deployed on a PC, with 16GB RAM and one Intel Dual-Core i7-6600U CPU.
    The main camera carries an OV5693 sensor and captures images with a size of $640\times480$ in pixels.
    18 volunteers participated in evaluating the accuracy of the selected methods.
    The heads of volunteers should be kept as still as possible in detection process.

    As described in Section \ref{sec:security}, video-based attacks are capable of imitating iris movements of real users. Thus, we mainly ponder the possibility for video attacks to spoof IriTrack.
    We assume that a potential adversary can learn typical parameters of IriTrack, such as angle type and segment length. To simulate these attacks, we record 50 different video clips (with random combinations of these parameters) for each of the 18 volunteers (i.e., 900 clips in total) with consistent indoor light intensity of 350lux.
    %To simulate an attack, we randomly select and replay a clip to spoof the targeted detection systems at each round of detection.
    We also test several scenarios with varying environmental factors to evaluate the flexibility of the proposed system against environmental changes.
    }

    \textbf{Summary of experimental results.}
      \begin{itemize}
        \item Among all potential combinations of parameter values, we find candidates that achieve a better balance between time overhead and detection accuracy, i.e., $s = 500$ and $L = \left \{ 150, 200, 250 \right\}$.
        Angles in 45deg and 90deg are hard for following so weights for these two kinds are relatively low.
        \item The average time overhead of liveness detection with IriTrack is roughly 3,845ms, which is dominated by the tracking module. IriTrack achieves higher detection accuracy in detecting 2D spoofing attacks, with an $\rm{F}_1$ score of 95.4\%.
        \item The probability-based random pattern generation model can reach a balance between processing time and detection accuracy.
        \item The performance of IriTrack can be maintained in a relatively stable and high level when environmental conditions change. Lowering circumstance brightness can help increase detection accuracy.
        %{\color{blue} And the system can keep its accuracy in a similar level as long as the distance between the tester and camera allows irises being located with steadiness.}
      \end{itemize}

  %Experiment Settings

  \subsection {Evaluation of Impacts of Parameters on Time and Accuracy} \label{sec:parameter}
    Now, we investigate how the time cost and accuracy vary according to different values of the parameters in IriTrack.

    As stated above, a slipper moving along the poly-line displayed on the screen is employed to help users to focus on the path and adjust the movement speed of their eyes.
    Thus, the time spent on iris tracking is positively correlated with the ratio of the total length of the given poly-line to the speed of the slipper.
    Intuitively, a shorter path with a faster slipper would significantly reduce the time interval for collecting trajectories.
    However, a fast-moving slipper may make users feel uncomfortable and also reduce the number of captured points, leading to a significant decrease of measurement accuracy.
    Therefore, we focus on trade-offs between time overhead and accuracy with varying parameter settings.
    \begin{table}[h]
      \caption{Parameters used for generating patterns}
      \label{tab:scale_parameters}
      \centering
      \renewcommand\arraystretch{1.3}
        \begin{tabular}{|c|c|c|c|c|c|}\hline
          \tabincell{c}{\# angle \\ types} & \tabincell{c}{\# line \\ lengths} & \tabincell{c}{\# speed \\ types} & \tabincell{c}{\# unique \\ patterns} & \tabincell{c}{\# tests \\ per pattern} & \tabincell{c}{\# total \\ tests} \\ \hline
          6 & 5 & 6 & 180 & 40 & 7,200 \\ \hline
        \end{tabular}
    \end{table}

    {\textbf{Dataset}.
    In order to clearly understand the impact of different parameters,
    the generated pattern is determined and simplified into a poly-line consisting of only 3 dots (i.e., 2 segments with a single angle).
    We assign the two segments with the same length, thus the total length of the line segments in a pattern, $l$,
    is twice the length of each segment.
    As summarized in Table \ref{tab:scale_parameters}, all combinations of parameters  $\theta$, $l$ and $s$ result in 180 unique patterns. Given a specific pattern, 4 volunteers are involved and each completes 10 times. The following figures show the average results of each pattern.}

  %  Now, we investigate the impact of $l$ and $s$ on time cost and accuracy.

    \begin{figure}[t]
      \begin{minipage}{0.48\textwidth}
        \centering
          \includegraphics[width=0.8\textwidth]{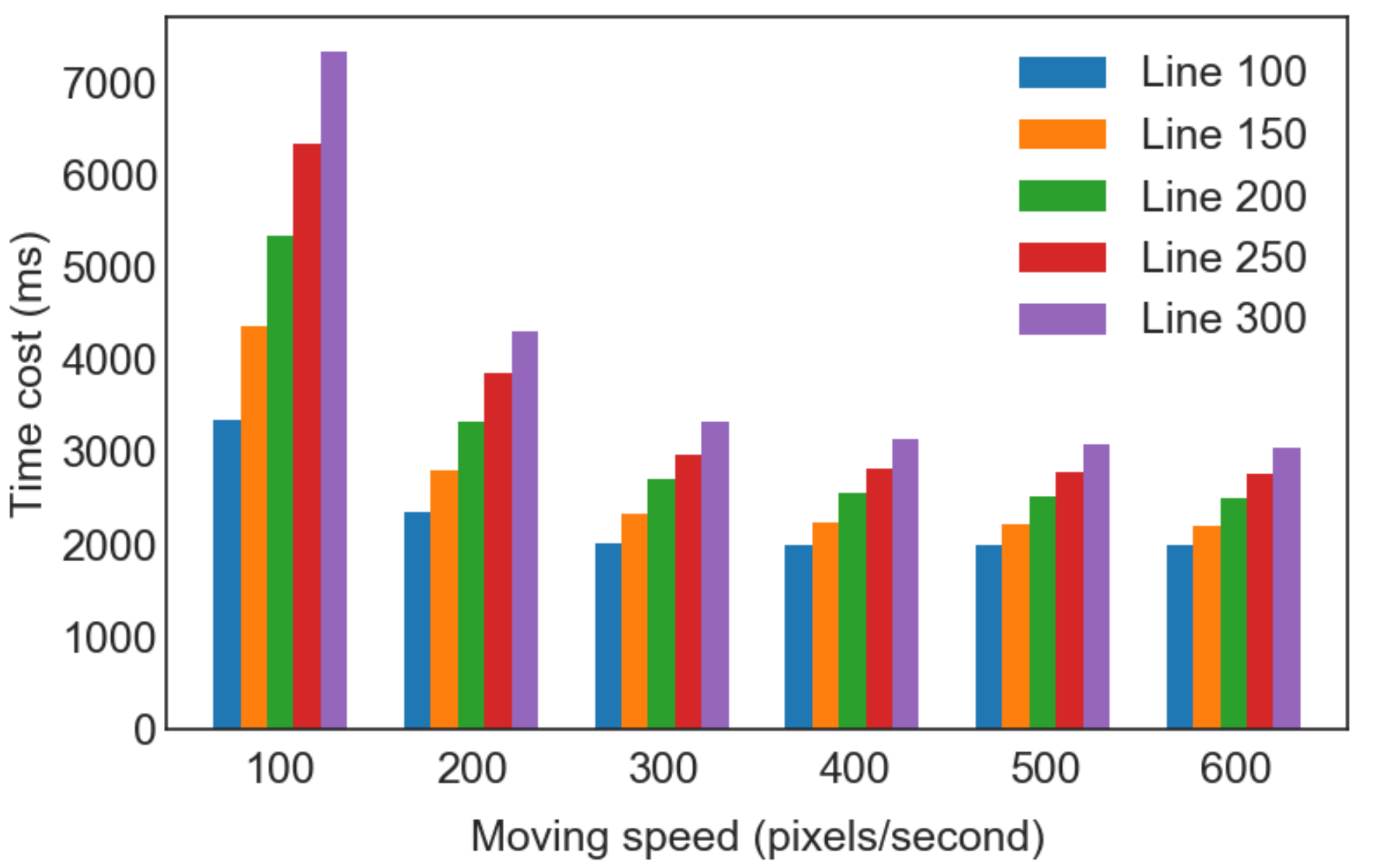}
        \end{minipage}
      \caption{Average time cost with various speeds and line lengths. For each combination, we calculate the average time cost of all kinds of angles in $\Gamma$.}
      \label{fig:timeoverhead}
    \end{figure}

    \begin{figure}[t]
      \begin{minipage}{0.48\textwidth}
        \centering
          \includegraphics[width=0.8\textwidth]{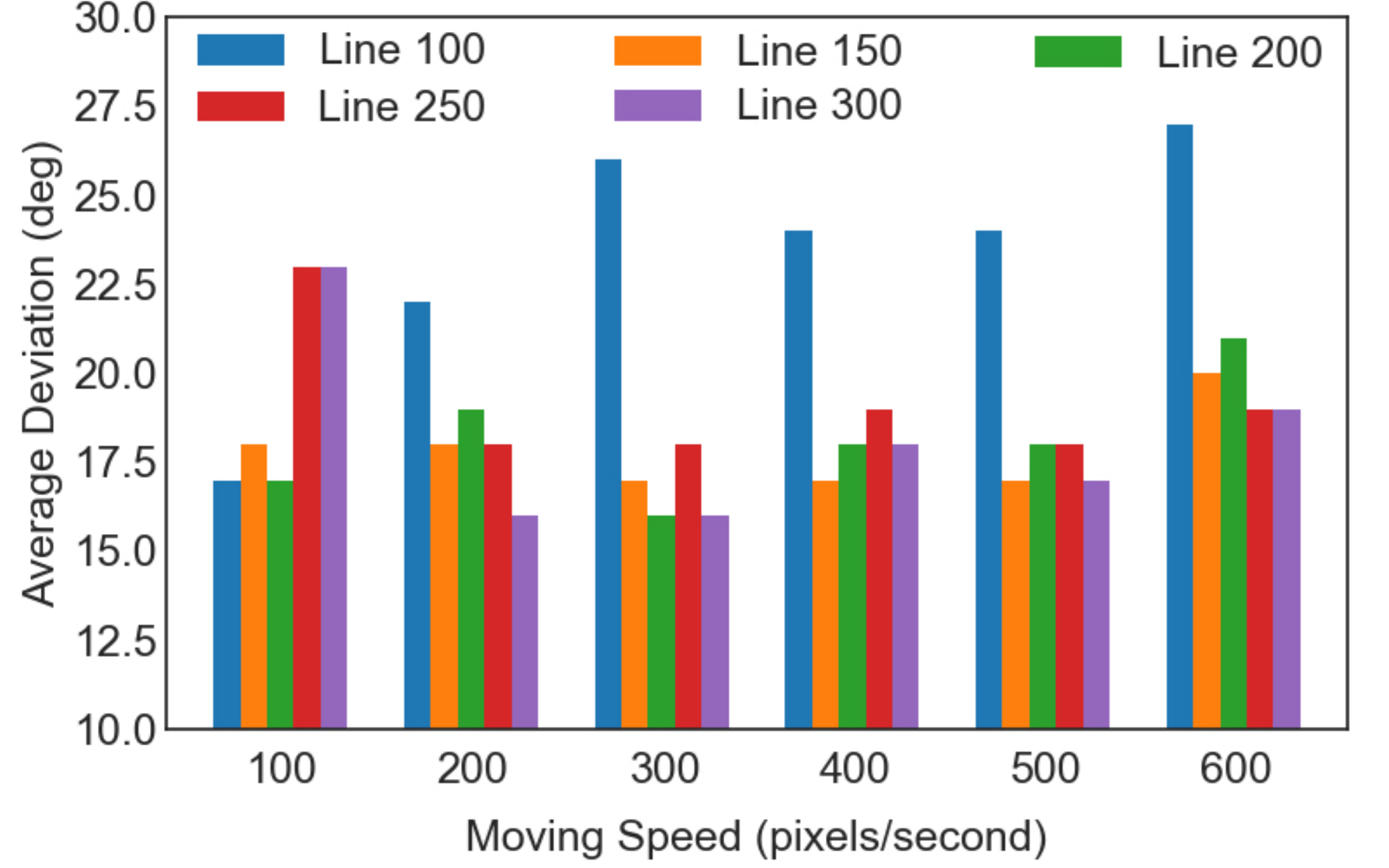}
        \end{minipage}
      \caption{The average matching deviation with varying speed $s$ and total length $l$. For each combination, values are derived from the average of measured deviation of all kinds of angles in $\Gamma$.}
      \label{fig:average_deviation_for_angles_b}
    \end{figure}

    \begin{figure}[t]
        \begin{minipage}{0.48\textwidth}
          \centering
          \includegraphics[width=0.8\textwidth]{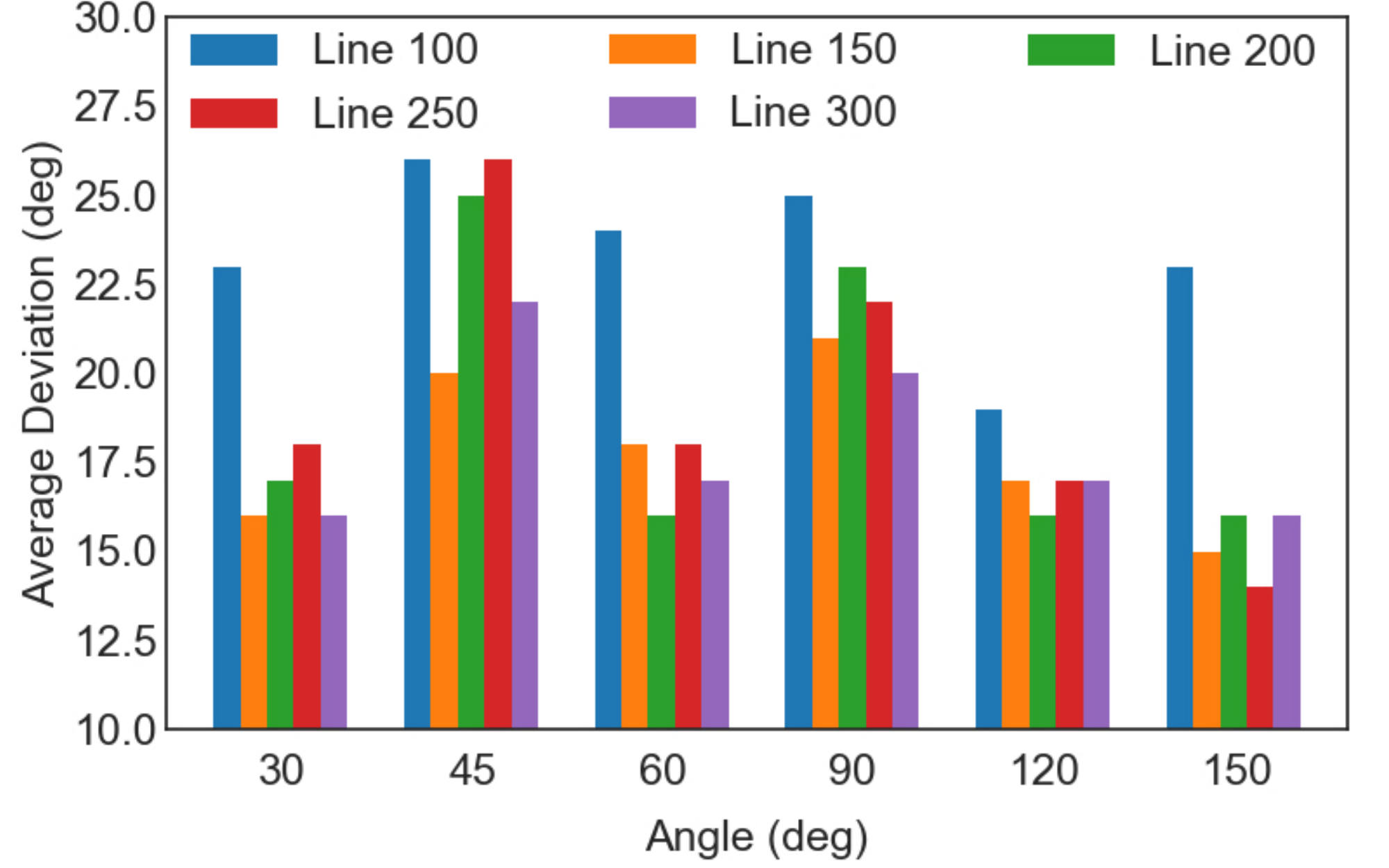}
        \end{minipage}
        \caption{The average matching deviation for different angles with various line lengths when $s=500$.}
        \label{fig:average_deviation_for_different_angles}
    \end{figure}

    The average time cost of the 6 typical kinds of angles with varying $s$ and $l$ are plotted in Fig. \ref{fig:timeoverhead}.
    We can find that at each fixed moving speed, the time spent on tracking grows as the line length increases.
    Thus, shorter lines contribute to a reduction in tracking time.
    When fixing the line length, speeding up the slipper's movement also reduces the time overhead for tracking.
    Thus, to achieve more efficiency in terms of time cost, combinations of shorter lines and higher speed are preferred.

    With the same settings as in Fig. \ref{fig:timeoverhead}, we exhibit the average matching deviation of angles with varying speeds and line lengths in Fig. \ref{fig:average_deviation_for_angles_b}.
    We observe two typical combinations leading to higher deviation, which are referred to as \emph{underspeed} and \emph{overspeed} cases.
    The underspeed cases happen when setting a low speed with relatively longer lines, e.g., the rightmost two bars at the speed of 100, as users would unconsciously try to predict the position of the slipper, making the tracking speed vary during the verification process.
    The overspeed cases happen when setting a high speed with relatively shorter lines, e.g., the length of 100 at a speed larger than 100.
    This is because shorter lines restrict eye movements with in a rather small area on the screen, making IriTrack more difficult to recover trajectory accurately.

    \textbf{Parameter selection result.} An appropriate combination of line length and moving speed leads to a better balance between accuracy and time efficiency. According to the results depicted in Figs. \ref{fig:timeoverhead} and \ref{fig:average_deviation_for_angles_b}, we set $s=500$ and $L=\left \{ 150, 200, 250 \right \}$ hereafter.

    Recall that each angle is associated with a weight, indicating the probability of it being selected when generating a pattern.
    Now, we describe the rationale for weight assignment.
    Among the 6 angles, the average of disparities between the measured and actual angles reaches a value of 20deg. However, considering angles of 45deg and 90deg which are harder for tracking, to ensure the performance for these 2 kinds, we set $C_0=25\rm{deg}$ which is their average deviation.

    For each of selected 6 kinds of angles, by calculating the frequency that a corresponding test case has a disparity no larger than $C_0$, we assign such frequencies as weights, as shown in Table \ref{tab:weights}.
    In Fig. \ref{fig:average_deviation_for_different_angles}, we demonstrate the matching deviation for each angle using the parameters recommended above.
    For instance, angles of 45deg and 90deg are relatively difficult for tracking and thereby their weights are lower than those of the rest angles.

    Based on the probabilities for generating different angles, $P_\Gamma(\theta)$, we utilize the highest, lowest, and average probabilities, to estimate the goodness of a pattern consisting of a certain number of angles. In general, a pattern should at least contain 2 angles with a corresponding goodness of 1.4.
    We take it as the baseline of goodness and set $G_0=1.4$.

    \begin{table}[h]
      \caption{Weights of Angles}
      \label{tab:weights}
      \centering
      \renewcommand\arraystretch{1.3}
      \begin{tabular}{|l|c|c|c|c|c|c|}\hline
        Angle & 30deg & 45deg & 60deg & 90deg & 120deg & 150deg \\  \hline
        Weight & 0.766 & 0.566 & 0.766 & 0.7 & 0.8 & 0.813 \\ \hline
      \end{tabular}
    \end{table}

  \subsection{Evaluation of Performance of IriTrack}

    Using the parameters determined above, we now evaluate the performance of IriTrack versus other counterparts.
    We employ the well-known accuracy criteria,
    i.e., precision (indicating the percentage of real faces detected in all instances detected as \emph{real}) and recall (indicating the percentage of real faces detected in all real faces in the ground-truth).
    Moreover, $F_1$ is calculated as $\frac{2 \cdot precision\cdot recall}{precision + recall}$.

    {
    \textbf{Dataset}.
    Each of the 18 volunteers is tested 40 times, which leads to a total number of $18\times40=720$ genuine cases.
    We also simulate 720 attack cases, which are conducted as follows:
    Considering that the pattern in IriTrack is generated randomly,
    we replay a clip selected randomly from the 900 clips to spoof the targeted detection systems at each round of detection.}

    \textbf{Time efficiency.}
    For the 4 selected face liveness detection systems, we record their average time costs for detection.
    OptFlowSys spends the most time as it requires the tester's head to swing slowly for several times while detecting the directional changes of optical flow.
    On the contrary, FlashSys needs the least time, as it captures and compares only two images in each round of detection, i.e., one without external light source and the other with flash turned on. However, the flash light is directly applied to the face of testers during each procedure, making the system less user-friendly.
    IriTrack holds a tolerable time cost, i.e. less than 4 seconds, which is comparable to that achieved by ncIriTrack.

    \textbf{Accuracy and security.}
    In order to reduce the influence of environmental factors, the selected methods are tested simultaneously.
    Besides tests with real persons, we present several instances of video attacks. The video attacks are conducted as follows:
    A series of video clips recording random iris movements are prepared in advance,
    and one video clip is randomly selected and displayed in front of the camera, attempting to cheat the liveness detection system.

    \begin{table}[t]
      \caption{Detection results of the tested detection systems}
      \label{tab:compare_two_real}
      \centering
      \renewcommand\arraystretch{1.3}
        \begin{tabular}{|l|c|c|c|c|c|}\hline
          System & Time (ms) & Precision & Recall & $\rm{F}_1$ score \\ \hline
          IriTrack & 3,845 & 95.2\% & 95.6\% & 95.4\% \\ \hline
          %rnIriTrack & 12,434 & 94.80\% & 95.18\% & 94.99\% \\ \hline
          ncIriTrack & 3,799 & 85.5\% & 80.4\% & 82.9\% \\ \hline
          FlashSys & 3,200 & 88.2\% & 92.4\% & 90.2\%  \\ \hline
          OptFlowSys & 7,350 & 77.6\% & 78.8\% & 78.2\% \\ \hline
        \end{tabular}
    \end{table}

    The detection accuracy of each system is presented in Table \ref{tab:compare_two_real}.
    IriTrack achieves the best performance in distinguishing between live real faces and fake faces.
    In IriTrack, patterns are generated with a random number of angles and lines, where the degree of each angle and the length of each line are also randomly selected from given sets. This greatly reduces the probability that a video attack successfully predicts a pattern.

    The fundamental goal of liveness detection is to be accurate, e.g., identifying more spoofing attacks in the ground truth, and avoiding false alarms.
    Thus, compared with FlashSys, one may prefer to use IriTrack for achieving higher accuracy with a slight increase of detection delay.

    {From the results collected at this stage,
    we pick 30 subsets, each of which contains detection results of randomly selected 50 cases (i.e., half with the genuine cases and half with the attack cases).
    Then, $F_1$ score of each subset is calculated.
    We find that for both FlashSys and OptFlowSys, there are statistically significant difference with 95\% confidence in comparison with IriTrack using the Student's t-test.}
  %the Student's t-test is applied, which reveals that there are statistically significant difference with 95\% confidence in comparison of the performance of IriTrack against both the performance of FlashSys and performance of OptFlowSys.}

    \begin{table}[h]
      \centering
      \caption{Liveness detection results for ncIriTrack and IriTrack}
      \label{tab:security}
      \centering
        \renewcommand\arraystretch{1.3}
        \begin{tabular}{|p{0.06\textwidth}|p{0.06\textwidth}<{\centering}|p{0.06\textwidth}<{\centering}|p{0.06\textwidth}<{\centering}|p{0.06\textwidth}<{\centering}|}\hline
          \multirow{2}{*}{Scenario} & \multicolumn{2}{p{0.12\textwidth}<{\centering}|}{ncIriTrack} & \multicolumn{2}{p{0.12\textwidth}<{\centering}|}{IriTrack} \\
          \cline{2-5} & Accepted & Rejected & Accepted & Rejected \\ \hline
          Real face & 80.4\% & 19.6\% & 95.6\% & 4.4\% \\ \hline
          Video & 13.6\% & 86.4\% & 4.8\% & 95.2\% \\ \hline
        \end{tabular}
    \end{table}

    The effectiveness of the timestamp-based optimization in IriTrack can be demonstrated by the comparison between ncIriTrack and IriTrack, as shown in Table \ref{tab:security}.
    With the help of timestamps, moving angles of irises can be more precisely recovered.
    Therefore, with IriTrack, more legitimate testers get passed (i.e., 95.6\% vs. 80.4\%) and more video attacks are successfully recognized (i.e., 95.2\% vs. 86.4\%).

    \textbf{Summary of performances.}
    The experiment results show that with IriTrack, the detection process takes less than 4 seconds and the $F_1$ score reaches 95.4\%.
    Thus, IriTrack owns the highest detection accuracy with a moderate time overhead.

  \subsection{Investigation of Pattern Scales}\label{sec:scale_pattern}
  The performance of IriTrack is largely determined by the generated patterns.
  This subsection investigates how the performance varies with patterns in different scales. We classify all the generated patterns according to the number of angles they contain.

  As stated in the last subsection, time and accuracy are crucial indicators of performance.
  Table \ref{tab:amount_angles_performance} reveals the results of the experiments, the last column indicates the detection accuracy when a pattern is tested by video spoofing attacks.
  The most complex pattern has the highest security and also the the highest time overhead.
  Generally, a pattern with more angles can certainly possess more line segments as well as break points, which results in growth of time consumption.
  Noticing that over all kinds of patterns, the security can be maintained in a relatively high level.

  Table \ref{tab:amount_angles_performance} also demonstrates that the probability-based model for pattern generation provides a flexible way to balance the tradeoffs between time efficiency and security.

    \begin{table}[h]
      \caption{Performance of patterns with different numbers of angles}
      \label{tab:amount_angles_performance}
      \centering
      \renewcommand\arraystretch{1.3}
        \begin{tabular}{|c|c|c|c|c|}\hline
          Angle Count & Tracking Time & Pattern Frequency & $F_1 score$ \\ \hline
          3 & 3,436ms & 55\% & 94.4\% \\ \hline
          4 & 3,634ms & 30\% & 95.3\% \\ \hline
          5 & 4,246ms & 11\% & 95.6\% \\ \hline
          6 & 4,875ms & 3\%  & 96.5\% \\ \hline
          7 & 5,217ms & 1\%  & 98.2\% \\ \hline
        \end{tabular}
    \end{table}

  \subsection{Evaluation of Environmental Impacts}
  This subsection evaluates the effect of the environment on the performance of IriTrack, including the light conditions and face-camera distances.
  %{\color{blue} The following experiments can be divided into 14 parts according to the environment settings, including 7 parts under different light intensities and other 7 parts under various face-camera distances. In each part, each participant is tested 5 times.}

  \begin{table}[h]
    \caption{Detection results with IriTrack when environmental  brightness varies}
    \label{tab:compare_brightness}
    \centering
    \renewcommand\arraystretch{1.3}
      \begin{tabular}{|c|c|c|}\hline
        \tabincell{c}{Intensity (lux)} & Intuitive description & $\rm{F}_1$ score \\ \hline
        1 & \tabincell{c}{Indoor, evening, screen light only} & 96.6\% \\ \hline
        25 & \tabincell{c}{Indoor, evening, with daylight lamp} & 95.6\% \\ \hline
        150 & \tabincell{c}{Indoor, afternoon, curtain closed} & 95.2\% \\ \hline
        350 & \tabincell{c}{Indoor, afternoon, natural light} & 95.4\% \\ \hline
        830 & \tabincell{c}{Indoor, afternoon, near a window} & 94.2\% \\ \hline
        2700 & \tabincell{c}{Outdoor, afternoon, cloudy} & 91.7\% \\ \hline
        10000 & \tabincell{c}{Outdoor, afternoon, sunny} & 91.6\% \\ \hline
      \end{tabular}
  \end{table}

  \textbf{Light Intensity.}
    In liveness detection systems, images of users are taken by cameras for further analysis.
    All previous experiments are conducted in a general indoor condition with a light intensity of 350lux.
    Next, we keep the brightness of the displaying screen at the same level (i.e. 250lux) and evaluate the performance of IriTrack by varying the environmental light intensities. {
    For video attacks, we keep using the same video dataset and replay strategy as mentioned earlier.
    Note that the device for replaying attack clips has a screen, which increases the environmental light intensity by 200lux on average.}

    The results are summarized in Table \ref{tab:compare_brightness}.
    We can find that the detection accuracy in terms of $F_1$ score maintains at a relatively high level as the environmental light intensity changes.
    An intensive sunlight slightly reduces the accuracy for detecting face regions, because the screen in such a condition can be comparatively darker, making it harder for the testers to keep focused.

 %   In this condition, the light coming from the screen can be of help in preventing the presented face and irises from being too dim to be detected. At the same time, a darker environment can contribute to helping testers concentrate on tracking with patterns.

    \begin{table}[h]
      \caption{Detection results of IriTrack with various distances between testers and the camera.}
      \label{tab:compare_distance}
      \centering
      \renewcommand\arraystretch{1.3}
        \begin{tabular}{|l|c|c|c|c|c|c|c|}\hline
          Distance (cm) & 20 & 22 & 24 & 28 & 32 & 34 & 36 \\ \hline
          Eye traceable & $\times$ & $\times$ & $\surd$ & $\surd$ & $\surd$ & $\times$ & $\times$ \\ \hline
          $\rm{F}_1$ score (\%) & 69.4 & 76.9 & 95.2 & 95.4 & 95.1 & 78.2 & 75.1 \\ \hline
        \end{tabular}
    \end{table}

    \textbf{Face-Camera Distance.}
    Distance between the face and camera will influence the size of faces in obtained images, e.g., a shorter distance helps get a larger face with more details of iris movements.
    Daugman's algorithm used in IriTrack searches irises with radiuses in a pre-defined range.
    That is, to make irises successfully and accurately detected, testers have to put their heads at a proper distance to the camera so that each of the captured irises could have an appropriate size for further detection.

    The results are exhibited in Table \ref{tab:compare_distance}. We can find that the detection accuracy in terms of $F_1$ score reaches a steady level,
    as long as the face-camera distance is appropriate where irises can be traceable.

\section{Discussion}\label{sec:discussion}
  %-----------------------------------------------------
  Being different from most existing liveness detection methods, IriTrack does not rely on direct analysis on images acquired by cameras, thus it needs no online or offline training of image classifiers for liveness detection.
  We have shown its effectiveness in the previous section.
  This section mainly discusses issues that might affect its performance in practice.

  \textbf{Compatibility on different devices}.
  Screens displaying the generated patterns may differ in their physical sizes (in terms of inches) and effective rendering sizes (in terms of pixels).
  A physically small screen may have a larger pixel density, which makes a line rendered visually shorter.
  To get a consistent displaying effect on different devices,
  the pixels per inch (PPI) parameter can be involved, which converts lengths in pixels into values in device-independent inches by simple multiplications.

  \textbf{Defense against advanced mask-based attacks}.
  As mentioned earlier in Section \ref{sec:security}, an adversary may use a mask which enables camera-detectable eye movement to spoof IriTrack.
  As a liveness detection system, IriTrack is only responsible for verifying if a user is alive, irrespective of the user is authorized or not.
  In general, existing liveness detection systems which take eye reaction (e.g., movement and blinking) as an alive sign are vulnerable to such advanced attacks.
  To defend against these attacks,
  static feature analysis approaches \cite{kim2012face,boulkenafet2016face} can be incorporated into IriTrack, since masks are different from real faces in textures.

  \textbf{Assumption on user concentration}.
  It is worth noticing that the heads of users should be kept as still as possible,
  as intensive jitters occur when recognizing face regions by OpenCV even though positions of a head are changed negligibly. Currently, IriTrack records the global positions of irises for each frame.
  In order to improve the steadiness of algorithms that locate face regions, the iris tracking module can use the relative positions between a face and the irises to identify the movement of irises. In this case, such an assumption will no longer be necessary.

  We leave these improvement attempts as future work.

\section{Conclusion}\label{sec:conclusion}
  %----------------------------------------------------- Conclusion
  In this paper, we proposed a face liveness detection system named IriTrack, which performs detection by comparing iris trajectories with randomly generated patterns.
  Each module in IriTrack does not require special hardware and is easy to implement on commercial devices.
  Extensive experimental results demonstrated the effectiveness of IriTrack in fending against video-based spoofing attacks.
  In future work, we will further improve the time efficiency and compatibility of the proposed system.
%Conclusion

%\bibliographystyle{unsrt}
%\bibliographystyle{abbrv}
\bibliographystyle{ieeetr}
\bibliography{main}

\begin{IEEEbiographynophoto}
%\begin{IEEEbiography}
%[{\includegraphics[width=0.8in,height=1in,clip,keepaspectratio]{bio/bio-shen.pdf}}]
{Meng Shen} received the B.Eng degree from Shandong University, Jinan, China in 2009, and the Ph.D degree from Tsinghua University, Beijing, China in 2014, both in computer science. Currently he serves in Beijing Institute of Technology, Beijing, China, as an assistant professor. His research interests include privacy protection of cloud-based services, network virtualization and traffic engineering.
He received the Best Paper Runner-Up Award at IEEE IPCCC 2014.
He is a member of the IEEE.
%\end{IEEEbiography}
\end{IEEEbiographynophoto}

\vspace{-50pt}

\begin{IEEEbiographynophoto}
%\begin{IEEEbiography}
%[{\includegraphics[width=0.8in,clip,keepaspectratio]{bio/bio-ma.pdf}}]
{Zelin Liao} received the B.Eng degree in computer science from Beijing Institute of Technology, Beijing, China in 2017.
Currently he is a master student in the School of Computer Science, Beijing Institute of Technology.
His research interest is secure face recognition.
%His research interests include Cloud Computing and Secure Searchable Encryption.
%\end{IEEEbiography}
\end{IEEEbiographynophoto}

\vspace{-50pt}
\begin{IEEEbiographynophoto}
%\begin{IEEEbiography}
%[{\includegraphics[width=0.8in,clip,keepaspectratio]{bio/bio-zhu.pdf}}]
{Liehuang Zhu} is a professor in the School of Computer Science, Beijing Institute of Technology.
He is selected into the Program for New Century Excellent Talents in University from Ministry of Education, P.R. China.
His research interests include Internet of Things, Cloud Computing Security, Internet and Mobile Security.
%\end{IEEEbiography}
\end{IEEEbiographynophoto}

\vspace{-50pt}

\begin{IEEEbiographynophoto}
%\begin{IEEEbiography}
%[{\includegraphics[width=0.8in,clip,keepaspectratio]{bio/bio-rashid.jpg}}]
{Rashid Mijumbi} received a PhD in telecommunications engineering from the Universitat Politecnica de Catalunya (UPC), Barcelona, Spain. He was a Post-Doctoral Researcher with the UPC and with the Telecommunications Software and Systems Group, Waterford, Ireland, where he participated in several Spanish national, European, and Irish National Research Projects. He is currently a Software Systems Reliability Engineer with Bell Labs CTO, Nokia, Dublin, Ireland. His current research focus is on various aspects of 5G, NFV and SDN systems. He received the 2016 IEEE Transactions Outstanding Reviewer Award recognizing outstanding contributions to the IEEE Transactions on Network and Service Management. He is a Member of IEEE.
%\end{IEEEbiography}
\end{IEEEbiographynophoto}

\vspace{-50pt}

\begin{IEEEbiographynophoto}
%\begin{IEEEbiography}
%[{\includegraphics[width=0.8in,clip,keepaspectratio]{bio/bio-du.jpg}}]
{Xiaojiang Du} is a tenured professor in the Department of Computer and Information Sciences at Temple University, Philadelphia, USA.
Dr. Du received his B.S. and M.S. degree in electrical engineering from Tsinghua University, Beijing, China in 1996 and 1998, respectively. He received his M.S. and Ph.D. degree in electrical engineering from the University of Maryland College Park in 2002 and 2003, respectively.
His research interests are wireless communications, wireless networks, security, and systems.
He has authored over 200 journal and conference papers in these areas, as well as a book published by Springer.
Dr. Du has been awarded more than \$5 million US dollars research grants from the US National Science Foundation (NSF), Army Research Office, Air Force, NASA, the State of Pennsylvania, and Amazon.
He won the best paper award at IEEE GLOBECOM 2014 and the best poster runner-up award at the ACM MobiHoc 2014.
He serves on the editorial boards of three international journals.
Dr. Du is a Senior Member of IEEE and a Life Member of ACM.
%\end{IEEEbiography}
\end{IEEEbiographynophoto}

\vspace{-50pt}

\begin{IEEEbiographynophoto}
%\begin{IEEEbiography}
%[{\includegraphics[width=0.8in,clip,keepaspectratio]{bio/bio-hu.jpg}}]
{Jiankun Hu} is a Professor at the School of Engineering and IT, University of New South Wales (UNSW) Canberra (also named UNSW at the Australian Defence Force Academy (UNSW@ADFA), Canberra, Australia).
He is the invited expert of Australia Attorney-Generals Office assisting the draft of Australia National Identity Management Policy.
Prof. Hu has served at the Panel of Mathematics, Information and Computing Sciences (MIC), ARC ERA (The Excellence in Research for Australia) Evaluation Committee 2012.
His research interest is in the field of cyber security covering intrusion detection, sensor key management, and biometrics authentication.
He has many publications in top venues including IEEE Transactions on Pattern Analysis and Machine Intelligence, IEEE Transactions on Computers, IEEE Transactions on Parallel and Distributed Systems (TPDS), IEEE Transactions on Information Forensics \& Security (TIFS), Pattern Recognition, and IEEE Transactions on Industrial Informatics.
He is the associate editor of the IEEE Transactions on Information Forensics and Security.
%\end{IEEEbiography}
\end{IEEEbiographynophoto}

\end{document}